\documentclass[12pt,preprint]{aastex}

\shorttitle{Tidal interaction and merger in CGs}
\shortauthors{Coziol \& Plauchu--Frayn}

\begin{document}

\title{Evidence for Tidal Interaction and Merger as the Origin of Galaxy Morphology Evolution in Compact Groups}

\author{R. Coziol}
\affil{Departamento de Astronom\'{\i}a, Universidad de Guanajuato\\
Apartado Postal 144, 36000 Guanajuato, Gto, M\'exico}
\email{rcoziol@astro.ugto.mx}
\author{I. Plauchu-Frayn}
\affil{Departamento de Astronom\'{\i}a, Universidad de Guanajuato\\
Apartado Postal 144, 36000 Guanajuato, Gto, M\'exico}
\email{plauchuf@astro.ugto.mx}

\begin{abstract}
We present the results of a morphological study based on NIR images
of 25 galaxies, with different levels of nuclear activity (star
formation or AGN), in 8 Compact Groups of Galaxies (CGs). We perform
independently two different analysis: a study of the deviations of
isophotal levels from pure ellipses and a study of morphological
asymmetries. The results yielded by the two analysis are highly
consistent. For the first time, it is possible to show that
deviations from pure ellipses are produced by inhomogeneous stellar
mass distributions related to galaxy interactions and mergers.

We find evidence of mass asymmetries in 74\% of the galaxies in our
sample. In 59\% of these cases, the asymmetries come in pairs, and
are consistent with tidal effects produced by the proximity of
companion galaxies. The symmetric galaxies are generally small in
size or mass, inactive, and have an early-type morphology. They may
have already lost their gas and least attached envelop of stars to
their more massive companions.

In 20\% of the galaxies we find evidence for cannibalism: a big
galaxy swallowing a smaller companion. In 36\% of the early-type
galaxies the color gradient is positive (blue nucleus) or flat.
Summing up these results, as much as 52\% of the galaxies in our
sample could show evidence of an on going or past mergers. Our
observations also suggest that galaxies in CGs merge more frequently
under ``dry'' conditions (that is, once they have lost most of their
gas).

The high frequency of interacting and merging galaxies observed in
our study is consistent with the bias of our sample towards CGs of
type B, which represents the most active phase in the evolution of
the groups. In these groups we also find a strong correlation
between asymmetries and nuclear activity in early-type galaxies.
This correlation allows us to identify tidal interactions and
mergers as the cause of galaxy morphology transformation in CGs.

\end{abstract}


\section{Introduction}
It is largely accepted that galaxy formation and evolution are
influenced, if not mainly driven, by interactions with their
environment. Interactions, in particular, are assumed to be
responsible for the phenomenon of segregation of morphologies with
radius observed in cluster of galaxies (Oemler 1974; Dressler 1980).
The exact nature of these interactions, however, is still largely
unknown. The problem is complicated by the large range of properties
presented by the environment of galaxies, but also because these
properties may equally evolve with time (for example, during the
progressive formation of large scale structures, where most galaxies
are embedded).

In cluster of galaxies, one possible mechanism for morphology
transformation is ram pressure stripping (Gunn \& Gott 1972).
According to this model, a late-type spiral galaxy, falling at high
velocity into a system having a high density of intergalactic gas,
feels a pressure that strips the gas off from its disk. The
remaining galaxy, which is now deficient in gas, takes an
earlier-type appearance. However, for such mechanism to work,
clusters of galaxies must already be rich in intergalactic gas, the
origin of which is still a debated matter (e.g. Valageas et al.
2003). It may also be difficult to understand how a late-type spiral
galaxy with a small bulge can transform into a larger bulge
early-type galaxy by simply losing the gas from its disk (Dressler
1980).

Avoiding these problems, it was proposed that gas rich spirals
falling at high velocity into a rich galaxy environment would
experience ``galaxy harassment'' (Moore et al. 1996, 1998).
According to this model, the disk of a late-type spiral suffers many
deformations from harassment, leading to some gas being rip off from
the disk, but mostly flowing toward the center of the galaxy, where
it could trigger multiple bursts of star formation, which will
contribute in increasing its bulge (and possibly also fueling and
AGN).

Although galaxy harassment could easily explain the formation of
dwarf elliptical and spheroidal galaxies in clusters (Moore et al.
1996), the formation of luminous elliptical and S0 galaxies may
require more dramatic events. One such event, which is credited to
create mostly elliptical galaxies, is merger (Toomre 1977; Merritt 1984).
Alternatively, massive inflows of gas, produced by tidal forces
originating from galaxy-galaxy interaction, or galaxy-cluster
interaction (that is, the galaxy interacting with the mass of the
whole cluster), may trigger intense starburst episodes in the center
of a galaxy (and also fuel an AGN), depleting its disk of gas and
increasing the mass of its bulge (Byrd \& Valtonen 1990; Mihos \&
Hernquist 1996; Hibbard \& Mihos 1995; Henriksen \& Byrd 1998;
Fujita 1998).

It is usually understood that merger and tidal interaction are
mechanisms that work more efficiently when the density of galaxies
is high and the velocity dispersion of the aggregate of galaxies is
low. These conditions are different than those favoring ram pressure
stripping and galaxy harassment. In fact, these conditions fit best
those encountered in Compact Groups of Galaxies (CGs; for a
definition of CGs see Hickson 1982 or Hickson et al. 1992). We
expect, consequently, to see more clearly the effects of such
processes in these systems.

Evidence for interaction of galaxies in CGs were found in
morphological studies (Mendes de Oliveira \& Hickson 1994) and in
analysis based on the distribution of neutral and molecular gas
(Leon et al. 1991; Verdes-Montenegro et al. 2001). The effects
expected from tidal interactions were also shown to be consistent
with the nature and level of activities (starburst and AGN)
encountered in CGs (Coziol et al. 1998a, 1998b, 2000, 2004). What
may still be missing, however, is a clear view of the final phase of
the evolutionary process of galaxies in CGs, and evidence that tidal
interactions and mergers produce the morphological transformations
predicted by the models.

According to the Hierarchical Galaxy Formation theory, the end
product of galaxy evolution through tidal interactions and mergers
should be massive early-type galaxies. Consistent with this
prediction, Hickson et al. (1988) have found the number of
early-type galaxies in CGs to be significantly higher than in the
field. More recently, Andernach \& Coziol (2006) have shown that CGs
isolated from large scale structures possess a clear excess of S0
galaxies compared to the field, which implies that their formation
must certainly be favored in these systems. Yet, evidence of
``recent'' mergers does not seem as obvious as expected. One such
evidence, in particular, should be elliptical galaxies with blue
nuclei. According to King's merger model (1977) the blue nucleus of
an elliptical galaxy is a product of massive stars migrating towards
the center of the galaxy by dynamical relaxation. Alternatively,
blue nuclei could also be the signs of recent bursts of star
formation. Indeed, there are many examples in the literature (both
observational and theoretical) connecting mergers to intense
starburst episodes. In CGs, therefore, such evidence should be
predominant. Surprisingly, Zepf et al. (1991) have determined the
fraction of ``blue'' elliptical galaxies in CGs to be only a few
percent. Taken at face value, this result suggests either mergers
are not as frequent as expected in these systems or past mergers (or
starbursts) are too old to be observed in the optical. However, it
seems difficult to reconcile the hypothesis of old ages for the
mergers, and consequently for the formation of the CGs, with the
numerous examples of on-going interactions encountered in CGs, as
described above.

As Coziol et al. (2004) have shown, it seems obvious now that the
dynamical behavior of CGs does not follow the predictions made by
the ``fast merger'' model (G\'omez-Flechoso \&
Dom\'{\i}nguez-Tenreiro 2001). For example, Hickson et al. (1988)
have found the number of early-type galaxies increasing with the
velocity dispersion of the group, instead of decreasing as predicted
by the model. Coziol et al. (2004) have found the same behavior
correlated to a low level of activity, suggesting that CGs with
higher velocity dispersions are more evolved. One possible way out
of this dynamical dilemma is to assume different formation times for
CGS (Coziol et al. (2004): if massive CGs (those with high velocity
dispersions) formed before those having lower masses (low velocity
dispersion systems), we would necessarily expect them to be more
evolved today.

In their spectral analysis of 91 galaxies in 27 CGs, taken from the
sample of Hickson (Hickson Compact Groups or HCGs: Hickson 1982),
Coziol et al. (2004) have defined a global activity index for the
groups, which they used, in combination with the average
morphological type of the galaxies in each group, to establish three
evolutionary types. The level of evolution of a group was proposed
to increase from type A (low velocity dispersion CGs, where galaxies
are active and late-type), to type B (intermediate velocity
dispersion CGs, where interactions and possibly mergers are on
going), to type C (high velocity dispersion CGs, where galaxies are
inactive and early-type). This classification offers a simple
explanation for the apparently low frequency of merger remnants
encountered in CGs: because mergers happen on average at earlier
time during the formation of the groups, observational traces of
these events must be difficult to distinguish in the optical. On the
other hand, it may still be possible to observe such traces in the
NIR, under the form of blue nucleus or stellar mass asymmetries.

In the present article, we present the results of a morphological
study of galaxies in CGs, based on NIR images (J and K'). Our goals
are to test the hypothesis of different formation time for CGs and
to determine the role played by tidal interactions and mergers on
the evolution of their member galaxies. For our analysis, we apply
independently two different methods: the fitting of elliptical
ellipses on the isophotal levels of the galaxies and a determination
of their levels of asymmetry. The plan for the article is the
following. In Section~2, we present the characteristics of our
sample of CGs. In Section~3, we describe our observations and the
reduction process. In Section~4, we explain the two methods used for
our analysis. The results are presented in Section~5, followed, in
Section~6, by a short discussion. Our conclusions are exposed in
Section~7.

\section{Selection and properties of the sample}

The sample for our NIR analysis consists of 8 HCGs (25 galaxies),
taken from the sample previously studied in spectroscopy by Coziol
et al. (2004). The physical properties of the groups are presented
in Table~1. The properties of the galaxies are given in Table~2.
Note that our sample is biased towards CGs of evolutionary type B.
This bias was not intentional, but reflects the fact that the NIR
observations were carried out before the results of the spectra
analysis were available.

The advantage of observing in the NIR is that light in this band is
assumed to trace the distribution of the oldest stellar population
of a galaxy and, consequently, to follow the luminous mass. In order
to verify this assumption we have compiled in Table~3 the NIR
magnitudes in J, H and K from 2MASS for most of the galaxies in our
sample. In Figure~1, we compare the colors of our galaxies with the
average color sequences traced by different stellar spectroscopic
classes, as given by Frogel et al. (1978).

The colors of HCG~56a are typical of Supergiant stars reddened by
extinction (Frogel et al. 1987), which is consistent with its
spectral classification as a starburst galaxy. The colors of HCG~56b
are typical of a power-law (Kotilainen \& Ward 1994), which is
consistent with its spectral classification as an AGN. The colors of
the rest of our galaxies in our sample are those typical of
elliptical and S0 galaxies (Sadler 1984; Bender et al. 1993). The
slightly blue $H-K$ colors observed in four of our galaxies can be
explained by close interactions (HCG~79b and HCG~98a), and traces of
younger stellar populations (HCG~88c and HCG~94a), as our analysis
will show. In general, we conclude that the NIR light in our
galaxies trace the distribution of the luminous mass.

\section{Observation and reduction}

The observations were carried out with the 2.12-m telescope of the
Observatorio Astron\'omico Nacional, located in the Sierra San Pedro
M\'artir in Baja California, M\'exico, during two different runs:
the first one from the 3 to the 6 of September 2001, the second one
from the 4 to the 6 of April 2002. The images were obtained using
the IR Camera CAMILA (Cruz-Gonz\'alez et al. 1994), which is based
on four NICMOS3 detectors with $256\times256$ pixels, sensitive from
1 to 2.5$\mu m$. This instrument includes a diaphragm and a wheel
with 12 filters, which are cooled to reduce the background radiation
level. The optical system, which consists of a mirror and a focal
reducer designed for the f/4.5 secondary, gives a $3.6\times3.6$ sq.
arcminutes field of view, corresponding to a plate scale of 0.85
arcsecond per pixel. During our observations, the nights were clear,
with an average seeing at the telescope of 1.1 and 0.9 arcseconds in
J and K', respectively (FWHM of a standard star used for focus).

For each galaxy, J ($\lambda = 1.275\ \mu m$) and K' ($\lambda =
2.125\ \mu m$) images were obtained following a standard procedure.
To avoid saturation of the detector, 5 ``coadd'' images in J
(respectively 10 in K') of 20 seconds (respectively 1 seconds in K')
were added together to form one image of 100 seconds (respectively
10 seconds in K'). The total integration time for each galaxy, as
indicated in Table~2, was reached by adding a sequence of images
taken in a mosaic way. Because of the relatively large size of the
detector compared to the dimension of the galaxies, this mosaic
technique allows to determine the sky background while continuously
integrating on the object.

For photometry calibration, 3 standard stars were observed: FS14,
FS18 and FS21. These stars were taken from the UKIRT fundamental
standard stars list, available on the WEB page of the
observatory\footnote{ http://www.astrossp.unam.mx/estandar/}. For
reduction purposes, dome flats were also obtained with a 60-watt
bulb, connected to a potentiometer that was used at low intensity.

The reduction and photometry calibration were done using standard
routines in IRAF. First, the images were trimmed, using IMCOPY, to
reduce the vignetting effect (CAMILLA at the time was showing a
strong vignetting effect on one of its quadrant). Using IMSURFIT
(with the option RESPONSE) a surface was then fitted on the combined
dome flats (using a spline3 function with order one on both
dimensions), obtaining a normalized flat field image. This
normalized flat field image was divided from each target image using
IMARITH. To eliminate the sky contribution, median sky images,
obtained from the combination of four adjacent images in the mosaic
series, were subtracted from each corresponding image in the
sequence. The $(x,y)$ coordinates of the target object in each image
were then measured using the circular aperture photometry option (a)
in IMEXAMINE. These positions were used to determine the maximum
size of the final combined image. All the images were then trimmed
(with IMCOPY) to produce frames with the same dimension and where
the object occupies approximately the same position. Before
combining the frames, the position of the object in each image was
measured once again (the same way as before) to estimate the shifts
(a fraction of pixel) needed to put it exactly at the same position
(using IMSHIFT). From the frames, a mean image was finally obtained
using IMCOMBINE. Note that, once the standard ``recipe'' is
established, the use of IRAF scripts allows reducing a whole series
of images at the same time, which accelerates significantly the
reduction process.

The instrument magnitudes for the standard stars were obtained using
the PHOT package in IRAF. Calibration equations were calculated by
fitting linear regressions on the observation, using the average
atmospheric extinction coefficients for San Pedro M\'artir, as
previously determined by Tapia et al. (1986). Due to the low number
of standard stars observed, the uncertainties on the calibrated
magnitudes are relatively high: $\pm0.04$ in J and $\pm0.07$ in K'.

\section{Description of the analysis methods}

We applied independently two different methods for our analysis. The
first one consisted in fitting elliptical isophotes over the image
of the galaxies. A study of the deviations from pure ellipses yields
information on the morphology and mass distribution of the different
stellar populations (depending on the filter used). A detailed
description of the method and how it is used to search for evidence
of past interactions and mergers in galaxies can be found in Barth
et al. (1995). Note that because we are observing in the NIR, the
results yielded by our analysis are easier to interpret in terms of
inhomogeneous stellar mass distributions. This is because our NIR
images are not affected by stellar formation, interstellar gas or
dust extinction (as we demonstrated in Section~2).

The second method consisted in determining the levels of asymmetry
of the galaxies (Abraham et al. 1994, 1996; Conselice 1997;
Conselice \& Bershady 2000). The application of the technique is
straightforward. First the center of the galaxy is determined. This
correspond to the point $(x, y)$ in our images ($x$ and $y$ being
real) where the intensity is maximum. The center is determined
within a circle of about two pixels, which is slightly bigger than
the average size of the PSF. Then the image is rotated by
$180^{\circ}$ around this center. After verifying that the location
of the maximum has not shifted (otherwise we correct for this
shift), we subtract the rotated image from the original one. In the
residual image the asymmetries appear under the form of excesses of
light (together with their negative images at $180^{\circ}$; see
Figures 30 to 35). The level of asymmetry within a circular aperture
of radius $R$ is calculated using the following formula (Conselice
1997):

\begin{equation}
A^2 \equiv \frac{\sum_{}^{R} 1/4 (I_0 - I_{180})^2}{\sum_{}^{R}
I_0^2}
\end{equation}

Where $I_0$ is the intensity in the original image and $I_{180}$ the
intensity in the rotated image. The term $1/4 = (1/2)^2$ is the
correct normalization factor. It yields a level of asymmetry between
0 (completely symmetric) and 1 (completely asymmetric).

Because the galaxies in our sample are not isolated, the value of
the maximum asymmetry ($A_{max}$) in our analysis is arbitrary, and
not representative of the level of symmetry of the galaxy itself.
The explanation is the following. As we integrate further out along
the radius to calculate $A$, the presence of companion galaxies
becomes more and more important. Therefore, we stop our analysis
using the largest radius possible that minimizes this contribution.
Because the asymmetry index is cumulative, its maximum is almost
always attained at the maximum radius of integration. But since this
radius is arbitrary, the maximum asymmetry is consequently also
arbitrary.

What indicates symmetry in our analysis is the smoothness of the
$A/A_{max}$ curve with radius (see Figures 30 to 35). Comparing with
the residual images ($I_0-I_{180}$), it is easy to verify that in
our analysis asymmetries appear as structures in the $A/A_{max}$
curve. The amplitudes of these structures are proportional to their
relative intensities. Note that because of the cumulative nature of
the index, asymmetries near the maximum radius may become more
difficult to detect in the $A/A_{max}$ curve. Especially if they
have low intensities or if other asymmetries are located directly
opposite to their positions. However, the presence of these
asymmetries can always be detected in the residual images.

Our asymmetry analysis applies to the inner part of the galaxies. It
is not sensible to intergroup light or sky gradients. Our method
also minimizes the possible contamination by foreground stars.
Usually, we were able to eliminate most of these stars in advance.
This was done by masking the stars (using Imedit in IRAF). Only in
two cases (HCG~88b and HCG~88c) this correction was not possible.
This happened because of the position of the contaminating
``object'' very near the center of the galaxy. Note that in one
case, HCG~88b, we are not convinced this object is really a star.

\section{Results}

In Figures~2 to 9, we present the J images, in negative, of the
eight CGs in our study. In each image we also identify the type of
nuclear activity shown by the galaxies. The order of presentation
follows the sequence of increasing evolution of the groups.

The first group, HCG~88 (Figure 2), is classified as type A and is
one of the most active in our sample. Only two (out of four)
galaxies were observed: HCG~88b a Seyfert 2 (Sy2) and HCG~88c a star
forming galaxy (SFG). This group is also one of the most spatially
extended in our sample (covering 8~arcminutes on the sky). This is
why the two galaxies are shown separately.

The presence of spiral arms, visible in both galaxies is consistent
with their late-type classification. These two galaxies show a
luminous protuberance very near their centers. These were
interpreted as stars in SDSS (based only on their colors; Young et
al. 2000). According to our study, however, we cannot discard the
possibility of a merging galaxy for HCG~88b (see our analysis
below).

The next five groups (HCG~37, 40, 56, 79 and 98) were all classified
as type B. In this type of CGs we can see many evidences of
interactions: extended, mostly deformed, envelopes around the
early-type dominant galaxies (HCG~37a, 40a, 56c and 79a), tidal
tails (HCG~40e, 56b and 79b) and possibly even common envelops
(HCG~56c with 56d, the whole group HCG~79 and HCG~98). These
structures are impressive, considering that what we see in our
images are mostly stars. The nuclear activity in these systems is a
mixture of AGN (Sy2 and LINER) and star formation, this last
activity being more frequent in galaxies having later-type
morphologies.

From the point of view of activity, the evidence for interactions
are missing in the two groups in our sample classified as type C
(HCG~74 and HCG~94). This probably has to do with the early-type
morphology of the galaxies and to the higher level of evolution
assumed for these groups. However, the presence of extended and
possibly common stellar envelops, but not homogeneous or symmetric,
suggests these systems are not in dynamical equilibrium (Pildis et
al. 1995; Nishiura 2000; Da Rocha \& Mendes de Oliveira 2005).

Because the results for the fitted ellipses and asymmetry analysis
in the J and K' band turned out to be similar (consistent with the
absence of dust extinction), we present only those obtained with the
J filter, which have higher signal to noise levels. Results for the
ellipse fitting analysis are shown in Figures~10 to 22. The radii
that include 50\% ($r_{50}$) and 90\% ($r_{90}$) of the light are
indicated by two arrows in the upper graph. Details of the
morphologies of each galaxy, under the form of photometrically
uncalibrated contour profiles, overlayed on the negative J images,
are presented in Figures~23 to 29. Finally, the results for the
asymmetry analysis are presented in Figures~30 to 35. For this last
analysis we show not only the variations of $A/A_{max}$ as a
function of the radius, but also the residual images
($I_0-I_{180}$), which we found extremely useful to detect
asymmetries and determine their possible origins. Note that a
typical asymmetry in these images has a signal to noise level
$S/N\sim10$.

In the following we discuss the results of our analysis group by
group, referring the reader to the corresponding images.

\subsection{CGs of Type A}

The surface brightness profiles of HCG~88b (Figure~10) show
significative structures. The most important one, visible at $r \sim
6''$ (which also produces a red bump in the color diagram), is
related to an object very near the center of the galaxy (also
visible in Figure~23a). This object is at the origin of the large
amplitude variations in ellipticity, position angle and in the
parameters $A_4$ and $B_4$, observed in the isophotal profiles.
Farther away than $r_{50}= 9.3''$, other variations of high
amplitudes appear: the $A_4$ terms increases up to a value of 4 and
changes from disky ($A_4>0$) to boxy ($A_4<0$). This galaxy becomes
blue toward its center.

The asymmetry analysis of HCG~88b (Figure 30a) is fully consistent
with the results of the ellipse fitting analysis. Within the radius
$r_{50}$, we observe a bright asymmetry near the center. In the
opposite direction, at $r_{50}=12''$, we see another excess of
light, partially covered by the negative image of the central
object. This second asymmetry, together with lower intensity
asymmetries produced by the spiral arms farther away, are at the
origin of the variations of the parameters observed at the same
radii in the isophotal profiles.

As we mention earlier, our first impression was that the bright
asymmetry observed in HCG~88b was due to a foreground star. However,
the presence of the second asymmetry suggests another possibility.
Indeed, we do not expect a foreground star to be connected with any
other asymmetry. It may be, therefore, that what we see is evidence
for a special kind of merger called ``cannibalism'': that is, a
massive galaxy ``absorbing'' or ``swallowing'' a smaller mass
companion. The second bright asymmetry would then be a tidal
structure produced by the intrusion of an external entity (which
could be the nucleus of a smaller mass galaxy) near its nucleus.

Alternatively, the second asymmetry may also be related to the bar
(as suggested by the morphological classification) or to the spiral
arms. However, the evidence of a bar in our images are not
compelling and one would need a very asymmetrical arm to explain
this asymmetry. Without further observations we cannot reject the
hypothesis of a merger for this galaxy.

The analysis for HCG~88c is very similar to that of HCG~88b. This
galaxy also shows a blue nucleus and large amplitude variations of
the parameters of the elliptical fits (Figure~10). The most
important of these variations is produced by an object within
$r_{50}$ (Figure~23b). The fainter variations, observed farther
away, can be produced by the spiral arms. This is consistent with
what we see in the residual image (Figure 30b).

Note that contrary to HCG~88b, we do not observe an excess of light
in the direction opposite the location of the brightest asymmetry,
which could be interpreted as a possible tidal feature. In this
case, therefore, the probability for this object to be a foreground
star is much higher. The only asymmetries observed in this galaxy
would thus be those produced by the spiral arms. There is,
consequently, no apparent evidence of interaction in our analysis
for this galaxy.

\subsection{CGs of Type B}

The first example of a CG of type B in our sample is HCG~37, which
is formed by two large galaxies and a smaller one. The dominant
galaxy, HCG~37a, is an elliptical (Figure~23c) with very smooth
surface brightness profiles (Figure 11). This galaxy is boxy on
almost all its extension. Our analysis shows that the isophotes
become more elliptical at large radius, with an axis increasingly
deviating from the semi major axis (variations of parameter $B_4$).
In the residual image (Figure~30c), these variations are connected
to a large arc-like asymmetry, which increases in intensity toward
the northwest. This huge excess of light points toward the two
companions galaxies (particularly HCG37b), suggesting a possible
gravitational influence. Within $r_{50}=7.7''$ we also distinguish
two faint structures appearing as two peaks in the $A/A_{max}$
diagram (first panel in Figure~30c).

The second dominant galaxy, HCG~37b, is a spiral seen almost edge on
(Figure~23d). This galaxy becomes blue toward its periphery (Figure
11). The isophotal profiles are very smooth, with almost constant
parameters and large errors. This is expected for an edge on spiral
galaxy, due to the orientation of the galaxy. Consistent with this
orientation, the asymmetry residual image shows an excess of light
toward the side directly facing us (Figure~30d). However, we also
note a slight excess of light pointing in the direction of HCG~37a
to the southeast. This could be interpreted as a tidal effect (in
response to what we see in HCG~37a) or it could be that the galaxy
is also oriented toward its more massive companion. Both
interpretations are consistent with tidal interaction effects.

The last galaxy in this group, HCG~37c, is classified as an
intermediate ellipsoid galaxy (S0a; Figure 24a). The isophotal
profiles are relatively smooth, with almost constant parameters
(Figure~12). The galaxy as a marginally (considering the error) blue
center. Except for the important contamination of light seen at
large radius, and due to HCG~37a, the asymmetry residual image
(Figure~31a) reveals no structures, which is fully consistent with
the smooth isophotal profiles.

The second group of type B in our sample is HCG~40. This is a very
interesting group, with many evidences for interactions (Figure~4).
The dominant galaxy, HCG40a, is elliptical (Figure~24b). This galaxy
possesses a marginally blue center (Figure~12). Although the surface
brightness profiles are smooth, the parameters of the fitted
ellipses show multiple, large amplitude variations, suggesting an
inhomogeneous stellar mass distribution. This is confirmed by the
asymmetry analysis. In Figure 31b, we observe two bright plume-like
asymmetries: one to the north, with maximum intensity at
$r=7''$ and the other slightly farther than $r=20''$.
These structures produce the depressions in the parameters $A_4$ and
variations in $B_4$ observed at the same radii in the isophotal
profiles.

The second galaxy of this group, HCG40b, is an S0 galaxy
(Figure~24c). The isophotal profiles (Figure~13) show large
amplitude variations, within and up to $r_{50}$. These variations
are related to a bright asymmetry observed to the north
(Figure~31c), and a weaker extension to the west. These asymmetries
also produce the plateau observed in the $A/A_{max}$ diagram. Note
that there is once again a possibility for the brightest asymmetry
to be a foreground star (this would be the third case). The extended
excess of light, on the other hand, seems consistent with an
interaction feature produced by the proximity of HCG~40c.

This galaxy, HCG~40c, is an edge on spiral (Figure~24d). As
expected, the isophotal profiles (Figure~13) are smooth with almost
constant parameters (and large errors). However, we distinguish in
the residual image (Figure~31d) two bright asymmetries. One is
located at $r_{50}=7.1''$ to the southeast, and the other a little
bit farther to the northwest. Only the first one is clearly visible
as a bump in the $A/A_{max}$ diagram (first panel in Figure~31d).
The bright asymmetry to the southeast is connected to a bridge of
matter between HCG~40c and HCG~40e. At the same distance, but in the
opposite direction, we also observe a fainter asymmetry. The
numerous asymmetries observed in this galaxy are consistent with
tidal structures produced by interactions (and possibly a bridge of
stars) with its three companions (HCG~40a, 40b and 40e).

The galaxy HCG~40d is a barred early-type spiral (Figure~25a), which
seems slightly separated from the others. In Figure~14, the presence
of the bar can be inferred from the ``plateau'' appearing slightly
beyond $r_{50}= 3.7''$ in the surface brightness profiles. In the
same region the ellipticity increases with the radius, the position
angle decreases and the parameters $A_4$ and $B_4$ show large
amplitude variations.

In the residual image (Figure~32a), an intense asymmetry appears at
$r_{50}$. This asymmetry produces the variations of the
isophotal parameters observed at the same radius in Figure~14. We
can also distinguish two asymmetries beyond $r=11''$, located at
the extremities of the disk. The excess of light to the east, just
below the asymmetry, is a residue from a subtracted star.

The fact that this galaxy is probably seen near edge on complicates
the analysis. Once more, the brighter asymmetry could be due to a
foreground star (the fourth case). However, the presence of other
accompanying asymmetries may also suggest cannibalism or a global
response to the proximity of the companions.

The interpretation seems clearer for the last galaxy in this group,
HCG~40e. This galaxy is directly connected to HCG~40c by a bridge of
stars, as visible in Figure~25b. To decrease the importance of this
structure, our analysis concentrates on the most inner part
($r<7''$). Within this limit, the isophotal profiles are smooth
(Figure~14), showing very few variations of the fitted ellipse
parameters. The galaxy is boxy and shows the tendency to become blue
toward its center. The residual image (Figure~32b) reveals no
important asymmetries within ($r_{50}=5''$), which is fully
consistent with the isophotal analysis. Beyond this radius, however,
we find two bright asymmetries, one located in the opposite
direction to the bridge of stars (northeast), the other more intense
toward the west. This last one could be an extension of the bridge.
The interpretation of these asymmetries as tidal structures is
obvious.

The next group, HCG~56, offers other examples of strongly
interacting galaxies. The dominant galaxy in this group, HCG~56a, is
a late-type spiral, seen edge on (Figure~25c). Our preliminary
analysis of its color (Section~2) suggests that our NIR images may
be affected by star formation and significant amount of dust
extinction. The effect of extinction could very well explain the
peculiar color profile (Figure~15), which suddenly turns red toward
the extremity of the disk.

The residual image (Figure~32c) shows the usual pattern for an edge
on spiral galaxy, with the side facing us being brighter. This
brighter face has a patchy appearance, though, which could be the
effect of dust extinction. We distinguish at least two peaks in the
$A/A_{max}$ diagram. One is located within $4''<r<9''$ and the
other start at $r\sim 14''$. These asymmetries produce the
variations of the parameters $A_4$ and $B_4$ at the same radii. Once
again, the side pointing toward the direction of the other group
members (to the north), seems more intense. The possible
interpretations are similar to those suggested for HCG~37a.

The second galaxy in this group, HCG~56b, is a barred, lenticular
galaxy (Figure~25d). Our preliminary analysis suggests our NIR
images may be affected by an AGN, also with significant dust
extinction (Section~2). This is consistent with the sudden
variations observed near $r_{50}=2.4''$ in the isophotal profiles.
Like for HCG40d, the bar produces a plateau in the surface
brightness profiles around $r_{90}$ (Figure~15). Out of the central
region, the isophotal profiles varied smoothly, the eccentricity
increasing with the radius and the galaxy changing from disky to
boxy.

The residual image (Figure~32d) reveals a bright asymmetry at
$r=10''$, opposite to the direction of the bridge of stars with
HCG~56c. This asymmetry produces the large amplitude variations of
the isophotal parameters observed at the same radius.

The S0 galaxy HCG~56c is connected by a bridge of stars to HCG~56b
(Figure~26a). The color profile (Figure~16) is constant over its
entire extension. This galaxy has circular isophotes within
$r_{50}=2''$ and elliptical ones farther away. The parameters $A_4$
vary little, while the parameters $B_4$ increases with the
ellipticity. The residual image (Figure~33b) reveals no asymmetries.
The sudden variation of the position angle observed in the isophotal
profiles probably marks the passage from a bulge to a disk.

The analysis for HCG~56d is very similar to the previous one. This
lenticular galaxy (Figure~26b) shows smooth, continuously changing
profiles, and very few variations of the parameters $A_4$ and $B_4$
(Figure~16). The residual image (Figure~33b) reveals no asymmetries.

Because of its extreme position in our image, only the isophotal
analysis for HCG~56e was possible. This small S0 galaxy (Figure~26c)
shows almost no variations in colors (Figure~17), very few
variations in ellipticity or of the parameters $A_4$ and $B_4$. Like
for HCG 56c, and 56d, we would not expect to see asymmetries in this
galaxy.

Our next group, HCG~79, contains three major members. The dominant
galaxy, HCG~79a, is an early-type spiral (Figure~26d). The surface
brightness in J (we do not have K' for the galaxies in this group)
is very smooth (Figure~17). We see small variations of ellipticity
and a shift by $18^\circ$ of the position angle. The parameter $A_4$
is mostly disky and shows small amplitude variations over its entire
extension. These variations are accompanied by similar variations of
the parameter $B_4$. The residual image (Figure~33c) reveals two
bright asymmetries. One is located within $r_{50}=6.3''$ to the
west, and the other farther away to the east. These asymmetries
produce the variations of the parameters $A_4$ and $B_4$ observed at
the same radii.

The second galaxy of this group, HCG~79b, is a lenticular galaxy
(Figure~27a). Its surface brightness profile shows a bump half way
between $r_{50}= 5.3''$ and $r_{90}$, which seems to influence the
values of the other parameters (Figure~18). The residual image for
this galaxy (Figure~33d) reveals a very bright asymmetry to the
west. This is consistent with the peak between $5''<r<10''$
 in the $A/A_{max}$ diagram and explains the variations
of the isophotal parameters observed at the same radii. This
asymmetry is opposite to the direction of the tidal tail and may
form an extension to the bridge of stars connecting HCG~79b to
HCG~79c.

The bridge of stars (Figure~27b) perturbs the isophotal profiles of
HCG~79c (Figure~18). This is confirmed by the residual image
(Figure~34a). In the opposite direction to the main asymmetry
produced by the bridge we may distinguish a fainter one, within
$r_{50}=6.1''$. However, since this asymmetry does not seem to show up
in the graph of $A/A_{max}$, which is quite smooth, we would rather
classify HCG~79c as symmetric.

The last group of type B in our sample is HCG~98. We have only
information on the two first members: HCG~98a and 98b. The analysis
of this group is complicated due to the presence of a foreground
star (the only one confirmed by spectroscopy) between HCG~98a and
98b (Figure~27c). This star was deleted from our image before
performing our analysis.

The dominant S0 galaxy in this group, HCG~98a, is classified as a
barred lenticular, although the bar is not obvious in our images
(Figure~27c). The two surface brightness profiles (Figure~19), which
look convex between $r_{50}$ and $r_{90}$, may trace the bar.
However, we find such evidence quite weak and cannot conclude
positively on the presence of such structure.

Within $r_{50}=5.8''$, we observe important variations of the
ellipticity and position angle. Both parameters, $A_4$ and $B_4$,
show multiple variations of low amplitudes. The residual image
(Figure~34b) reveals two major asymmetries. One located within
$r_{50}$, the other farther away to the west, and associated to a
faint extension to the south. These asymmetries clearly produce the
variations of the isophotal parameters. These structures (in
particular the most central one) may reflect some advanced merger
phase, or correspond to tidal features, resulting from the
interaction with HCG~98b.

For the second galaxy, HCG~98b, we observe (Figure~19) a sudden
change of ellipticity and position angle at $r_{50}=2.9''$ and
multiple low amplitude variations of the parameters $A_4$ and $B_4$.
The residual image (Figure~34c) reveals an asymmetry located at
$r_{50} $ to the east, which produces the variations observed in
the isophotal profiles. We also distinguish an excess of light in
the direction of HCG~98a. Once again (for the fifth time), the
bright object at $r_{50} =1$ could be a foreground star. The
excess of light, on the other hand, seems related to a possible
bridge of stars, which would connect HCG~98b to HCG~98a (Figure~7
and Figure~27d).

\subsection{CGs of type C}

The first group of type C in our sample, HCG~74, is formed by three
early-type galaxies, embedded in an apparent common envelop of stars
(Figure~8). The dominant galaxy, HCG~74a, is an elliptical with a
light protuberance to the west (Figure~28a). This object affects all
the isophotal profiles (Figure~20).

In the residual image (Figure~34d), the protuberance corresponds to
a bright asymmetry located at $r =5''$, with some weaker
extensions passing through the center of the galaxy. In opposite
direction to this bright feature, to the east, we distinguish a much
more diffuse and extended asymmetry. The presence of this pair of
asymmetries is an argument in favor of cannibalism (and not a
contamination by a foreground star, which would be the sixth
example). Supporting this interpretation, we note that the excess of
light to the east does not point toward HCG~74b or HCG~74c.

The two companions, HCG~74b and HCG~74c (Figures 28b and 28c), show
similar isophotal profiles, with very few variations (Figures 20 and
21). Their residual images (Figure~35a and 35b) reveal no
asymmetries. Apparently, these galaxies are symmetric.

The second group of type C in our sample, HCG~94, is very similar to
the previous one (Figure~9), with the exception that the two
dominant galaxies in this group seem now comparable in size and mass
(based on comparable luminosities in the optical and NIR). These two
galaxies (Figures 28d and 29) have a blue nucleus (Figure~21 and
22). The isophotal profiles show only small amplitude variations of
the fitted ellipse parameters. The residual image of HCG~94a (Figure
35c) reveals one asymmetry just slightly farther than $r_{50} =
3.6''$ and a weak excess of light forming an arc-like structure
from east to north. A similar weak excess of light is observed in
HCG~94b (Figure~35d). This arc-like structure extends from west to
north. In HCG~94a the brighter asymmetry is consistent with
cannibalism, while the two arc-like structures could be tidal
structures produced by the interaction of the two galaxies.

\section{Discussion}

The results produced by our two methods, applied independently, are
highly consistent. This is of major importance for our analysis.
From a detailed comparison of the results, for the first time we can
verify that the variations of parameters of fitted ellipses in
galaxies reflect inhomogeneous stellar mass distributions, produced
by tidal interactions and mergers (Di Tulio 1979; Kormendy 1982;
Zaritsky \& Lo 1986; Bender \& M\"ollenhoff 1987; Barth et al.
1995). The two methods being applied independently, the consistency
of the results also puts heavy weights in our observations.

In Table~4, we summarize the results obtained from our asymmetry
analysis. A large fraction (74\% or 17 out of 23) of the galaxies in
our sample present evidence of asymmetries related to interactions
(we do not count HCG~88c, where the asymmetries are probably related
to the spiral arms). Of highly importance for the physical
interpretation of our result, in 59\% (10 out of 17) of the
asymmetric cases the asymmetries come in pairs. This is as expected
if they are the results of tidal forces.

In the seven galaxies where the asymmetry is not in a pair, the
excess of light either clearly points toward the projected position
of the companions (HCG~37a, 37b, 40b, 56a, 98b), or seems to form
part of a common envelop (HCG~94a and 94b). In general, therefore,
detected asymmetries in CG galaxies are directly connected to
interaction effects. This result confirms the analysis previously
made by Mendes de Oliveira and Hickson (1994).

For the tidal cases, we identify two possible origins for the
asymmetries. In five galaxies (HCG40a, 40c, 40e, 56b and 79b) one
asymmetry forms a pair with a bridge or a tail. In the other five
galaxies (HCG~88b, 40d, 79a, 98a and 74a), the brightest asymmetry
is accompanied by a weaker or more extended one, directly opposite
to its position. This we interpret as evidence for a special case of
merger, that we call cannibalism: a massive galaxy swallowing a
smaller mass companion.

Note that although we cannot eliminate the possibility of
contamination by foreground stars to explain the most intense
asymmetry in the candidate cannibal galaxies, our interpretation has
the advantage to yield a simple explanation for the presence of the
second accompanying asymmetry, whose origin, otherwise, would stay
undefined. This is the case of HCG~74a, for example, where the
excess of light directly opposing the brighter asymmetry does not
point toward neither of its two companion galaxies, and consequently
cannot be a tidal response to their proximity.

Our conviction in favor of the cannibalism interpretation stands on
three points. First, the frequency of cases where a star would
coincides with the nucleus of a galaxy in our sample looks
suspiciously high (24\%, 6 out of 25). This is considering that the
line of sight of the CGs in our sample are not passing through
particularly dense stellar regions. Second, we do not expect a
bright star to be apparently connected with another asymmetry. In
HCG 98, where we are sure the bright object is a star, and in other
cases where we have subtracted the stars before doing our analysis,
we find no relation between the detected asymmetries and the residue
left by the star subtraction. Finally, we do find numerous examples
of cannibalism in the local group (e. g. Martin et al. 2004;
McConnachie \& Irwin; Belokurov et al. 2006; Zucker et al. 2006;
Martin et al. 2006), and we expect consequently such phenomenon to
be even more frequent in CGs.

Another evidence in favor of past mergers in our sample is the high
frequency observed of early-type galaxies showing a flat or
positive color gradient (Zepf et al. 1991; Michard 1999; La Barbera
et al. 2003; Ko \& IM 2005). We count nine galaxies with such a
feature in our sample: HCG~37a, 37c, 40a, 40b, 56e, 94a, 94b, 98a
and 98b. These galaxies are mostly inactive (some activity is
observed in HCG~37a 37c and 56e, but only at a low level) and
asymmetric (except for HCG~37c and 56e, for which the information is
missing).

If we consider only the elliptical galaxies, 3 out of 6 are found to
be blue in their inner part: HCG40a, 94a, 94b. HCG94a and HCG94b
were also reported to be blue in the optical by Pildis et al.
(1995). HCG40a, which is marginally blue in our image, was not
reported to be blue by Zepf et al. (1991). However, these authors do
not present the color profile, which is obviously flat in our image.
Note that we do not find the color gradient of HCG37a presented by
Zepf et al. (1991) to be that blue either (positive slope), although
we concur on the flat color profile.

Note that we do not use the level of boxyness as evidence for merger
in our study for two reasons. First, this criterion usually apply to
elliptical galaxies, assumed to be in a final stage of evolution.
Our data suggest a much more complicated picture, the nature
boxyness or diskyness of the profiles frequently varying with the
radius. This behavior is consistent with the high number of
asymmetries observed in our galaxies, which suggest the early-type
galaxies in our sample are not in a final stage of evolution.
Second, recent simulations have shown that even for normal
elliptical galaxies the relation between boxyness and mergers is not
as straightforward as previously thought (see Naab et al. 2006a;
2006b), depending very much on the details of the mergers.

Summing up the galaxies with evidence for cannibalism to the
peculiar early-type galaxies, as much as 52\% of the galaxies in our
sample could thus show evidence for past mergers. This is much
higher (6\%) than what was found by Mendes de Oliveira and Hickson
(1994), in their morphology analysis based on the optical, or Zepf
et al. (1991) in their search for optically blue elliptical
galaxies. The frequency of merger remnants we have detected is now
much more consistent with what was originally expected in CGs.

Taken at face value, our results on past merger in CGs may seem in
contradiction with those found by Zepf et al. (1991) or Mendes de
Oliveira and Hickson (1994). However, we have to note that our
sample is biased towards CGs of type B, which according to the
scenario of evolution proposed by Coziol et al. (2004) represent the
most active phase in the evolution of the group. One would obviously
expect evidence for past mergers to be more prominent in these
systems. Our results, therefore, seems to confirm our classification
scheme based on activity.

For the six symmetric galaxies in our sample, we cannot eliminate
the possibility of asymmetries at a lower intensity level and higher
spatial resolution than reached in our analysis. However, it seems
obvious that the level of perturbation in these galaxies must be
significantly low compared to that in the other galaxies. These
galaxies turn out to be smaller in size (and probably in mass, as
judged from their NIR luminosity) than their companions. They are
also mostly inactive (the exception is HCG 56d), and have an
early-type morphology. Therefore, it could be that, having smaller
masses, these galaxies have already lost most of their envelop of
gas and least attached stellar disks to their more massive
companions. What is left, would be the components that are more
tightly bounded by gravitation to the galaxy, explaining the
apparent absence of asymmetries. This description would also fit the
entities found near the nuclei in the cannibal candidates.

From the point of view of nuclear activity, we cannot establish a
one to one relation with interaction. Although 77\% (10 out of 13)
of the galaxies showing some sort of activity in our sample also
show asymmetries related to interactions (again, we do not consider
HCG~88c), these cases represent only 55\% of the galaxies with
asymmetries. Neither the morphological types of the active galaxies
can explain the results: only 50\% of the active galaxies in our
sample have a late-type morphology.

However, if we consider the level of evolution of the groups, a
clear trend appears. In Table~5, we separated the groups according
to their evolutionary type. In the cases of type A and C our data
are too scarce to establish any firm tendency. But for the groups in
type B, we can clearly distinguish a connection between interaction
and activity: most of the galaxies that show asymmetries related to
tidal interactions and mergers are active, while only five have a
late type morphology. The fact that a high number of early-type
galaxies are active and perturbed in these systems directly connects
tidal interactions and mergers to morphological transformations.

Our observations support the morphological transformations of
late-type galaxies into earlier types in CGs. It remains, however,
to determine what could be the typical time scale for such
transformations. Recent merging models of gas poor galaxies suggest
tidal asymmetries in the NIR should be observable over a time scale
of the order of $0.4\pm0.2$ Gyrs (Van Dokkum 2005). Toledo et al.
(2006) found comparable merger time scales in their asymmetry
analysis of 42 E/S0 interacting pairs. On the other hand, Coziol et
al. (1998b) have shown that the usual indices for post-starburts in
CGs suggest ages higher than 2 Gyrs for the last bursts in the
early-type galaxies. Therefore, their seems to be a significant
difference between the two time scales. Note that Zefp \& Whitmore
(1991) already encountered a similar discrepancy comparing the
frequency expected of ``blue" elliptical galaxies with the dynamical
time scale for the evolution of the groups (the difference is of the
order of what we reported above). More generally, Schweizer \&
Seitzer (1992) found a similar problem in their study on the origin
of elliptical galaxies by mergers.

The hypothesis putted forward to explain the symmetric and
cannibalized galaxies may offer a new solution. If mergers in CGs
take place under dry conditions, that is, happening after most of
the gas has already been ejected, consumed into stars or burned by
an AGN, then, without the possibility of extra star formation, the
merging remnants would not be expected to be particularly blue (Naab
et al. 2006a, 2006b; Van Dokkum 2005). This would give the
impression that mergers in CGs are rare and post-starbursts older
than they really are.

One way to test this hypothesis is to determine the colors of tidal
features in dense environment. In absence of numerous massive stars,
these features would not be expected to be particularly blue. In
their analysis of galaxy interactions in clusters, Conselice \&
Gallagher (1999) have shown that tidal structures have a range of
colors extending from blue ($B-R \approx 0$) to red ($B-R \approx
2.5$). Comparing these colors to those of normal S0 and Elliptcal
galaxies (Michard 2000), one can see that these structures are
rarely blue (average of 1.46). In table~3, the galaxies in our
sample have an average color ($B-R$) of 1.63, which agree relatively
well with the distribution of the colors of the tidal structures
encountered by Conselice \& Gallagher. These authors suggest what
they observe is the result of interactions of the galaxies with the
cluster potential. This is also what we propose as the cause of dry
merger in CGs.

Other observations consistent with the hypothesis of dry merger in
CGs are the following: 1) the absence of antenna like merging
systems or luminous AGNs (Coziol et al. 1998a); 2) the fact that
star formation is generally low (Allam et al. 1999; Coziol et al.
2000), 3) the detection of galaxies which are deficient in gas
(Hutchmeier 1997), 4) the effects expected from truncation of star
formation on the evolution of stellar populations (Caon et al. 1994;
; Coziol et al. 1998b; de La Rosa et al. 2006). Finally, the
characteristics of the galaxies identified as candidates for dry
merger in our sample fit relatively well the definition given of
these objects by Bell et al. (2006). The highest prediction made by
these authors on the frequency of merger events expected in dense
environments are also in good agreement with the frequency of dry
merging candidate galaxies we encountered in our sample.

Putting everything together, the scenario that fit better our
observations in CGs of type B seem to be the following. As galaxies
join to form a group, the effect of tidal interaction with the
potential well of the whole system affect first their gas content:
part of the gas passes to the nucleus starting star formation and/or
refueling an AGN in gas, and part must go to the intergalactic
medium. Starting with spiral galaxies, this process would be
expected to produce mostly lenticular galaxies. These galaxies would
then start interacting with each other, and at one point merge
together, to form one giant elliptical galaxy. Under dry merger
conditions, the time scale of the process must be shorter than 2
Gyrs. This implies that CGs of type B formed relatively recently.

\section{Conclusions}

The two analysis applied independently are highly consistent. Both
methods suggest the galaxies in CGs are clearly not in equilibrium,
but in a process of transformation. The asymmetry analysis allows us
to identify the mechanisms behind this process: the galaxies
experience the effects of tidal interactions and possibly multiple
mergers.

Concerning tidal interactions, it is not clear what distinguish
galaxy-galaxy interactions from galaxy-group interactions. The
scenario we propose for the evolution of the groups suggests the
difference between the two mechanisms lies in the response of the
gas: through galaxy-group interaction, the galaxies would first
lose their envelop of gas and would then start interacting and
merging with each other under dry conditions.

Combining our observations on asymmetries with information on
nuclear activity we also have a definite view about the type of
transformations that is taking place in CGs: late-type galaxies
transform into earlier types (S0 and elliptical). From our analysis,
therefore, we clearly identify tidal interaction and merger as the
mechanisms responsible for the transformation of morphology of
galaxies in CGs.

As for the global evolution of CGs, our sample is unfortunately
still incomplete. We need more information on CGs of type A and C in
order to be able to test the hypothesis of different formation ages
for the CGs. Based on our data alone, however, the possibility that
all the galaxies in one CG will merge to form one giant elliptical
galaxy cannot be rejected.

This is where the hypothesis of a different formation ages for CGs,
depending on their mass, becomes interesting. The first CGs to reach
complete evolution would have been located in massive and dense
environments. Under such conditions, different groups could have
merge with each other to form larger structures (Andernach \& Coziol
2006). Consequently, we would not expect to see ``isolated"
elliptical galaxies as the product of CGs evolution, since these
remnants today would be in larger groups or cluster of galaxies
(Yoshioka et al. 2004).

Assuming merger of CGs were more frequent in the past, such events
could have played an important role in the formation of large scale
structures, like cluster of galaxies (Mihos 2004). Consequently,
many galaxies found today in clusters may have first evolved in
groups (Ellingson 2003; Mihos 2004). If this is the case, our
results would support the idea that the segregation of galaxy
morphologies observed in clusters today is mostly a product of tidal
interactions and mergers.

\acknowledgments We thank the CATT of San Pedro M\'artir for the
observing time given on the 2.12m telescope to realize this project
and all the personal of the observatory for their support. We also
thank Dr. Ascensi\'on del Olmo Orozco and Dr. Heinz Andernach for
discussing parts of this article with us. An anonymous referee is
also acknowledge for comments that have helped to improve the
clarity and quality of our presentation. Finally, we acknowledge the
support of CONACyT under grant 47282-F. This research has made use
of the NASA/IPAC Extragalactic Database (NED), which is operated for
NASA by the Jet Propulsion Laboratory, California Institute of
Technology.

\clearpage

\begin{figure}
\epsscale{.80}
\plotone{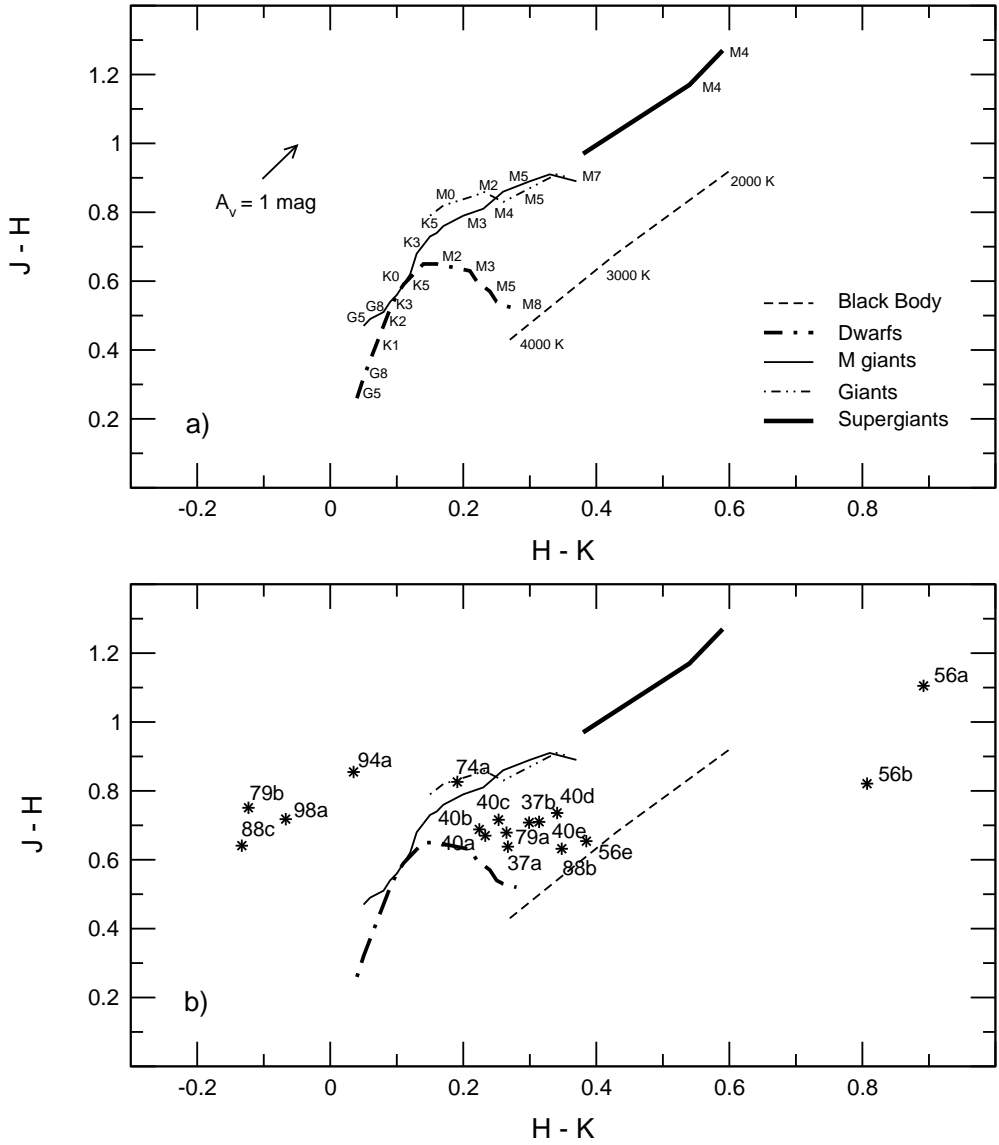}
\caption{a) Color-color diagram, using 2MASS, of
stars having different spectral types. b) The colors of the galaxies
in our sample are compared to those of the stars. Except for HCG~56a
and HCG~56b, the comparison shows that the colors of the galaxies
can be explain by a mixture of M dwarf and M and K giant
stars.\label{fig1}}
\end{figure}
\clearpage

\begin{figure}
\plotone{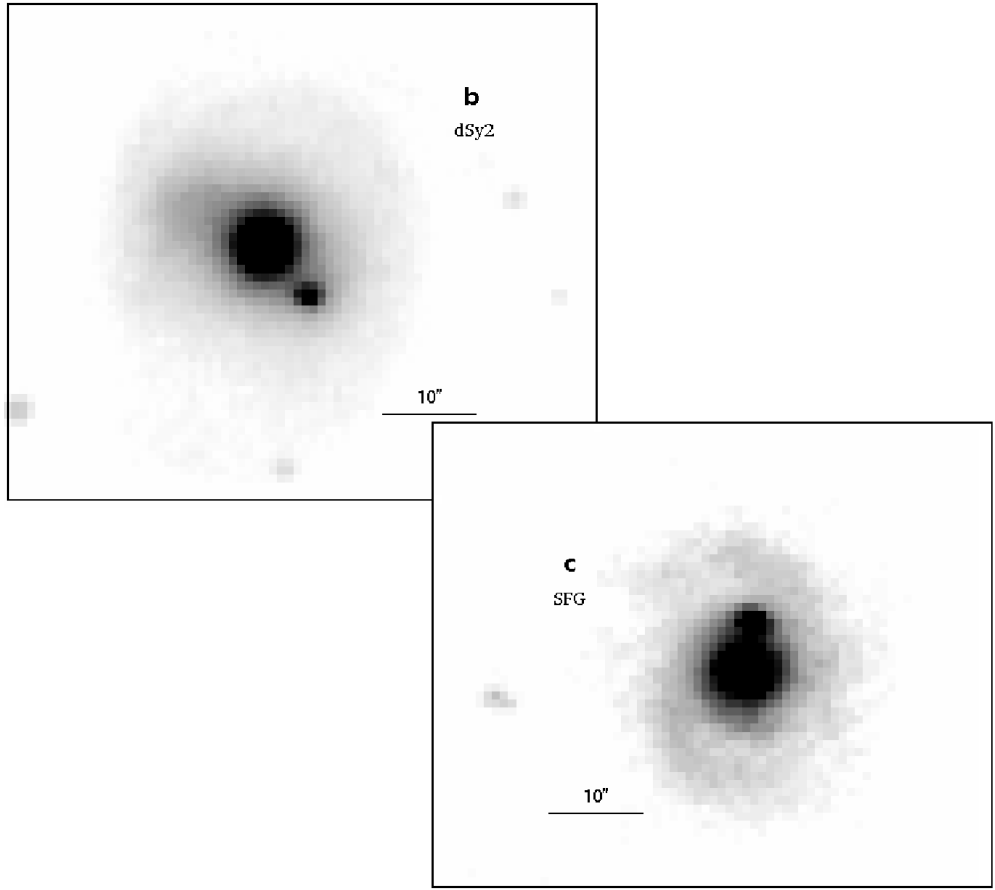}
\caption{Images in J band (in negative) of HCG~88b
and HCG88c. The north is up and east is to the left.
The fainter grey structures shown in these images have a S/N of 29 and 17, respectively in HCG 88b and HCG88c.
\label{fig2}}
\end{figure}
\clearpage

\begin{figure}
\plotone{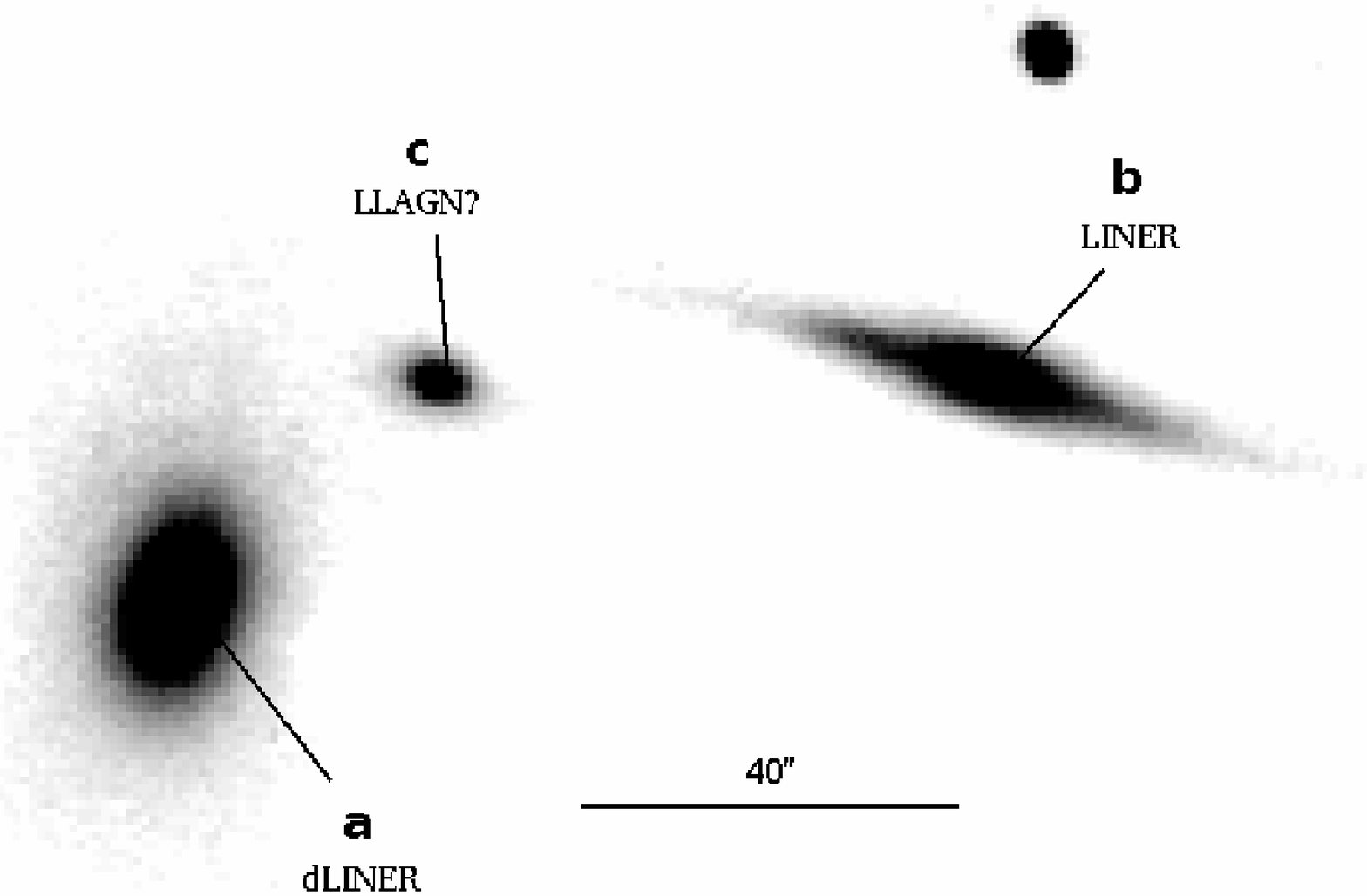} \caption{Same as Figure~2 for HCG 37.
The fainter grey structures shown in this image have a mean S/N of 13.
\label{fig3}}
\end{figure}
\clearpage

\begin{figure}
\plotone{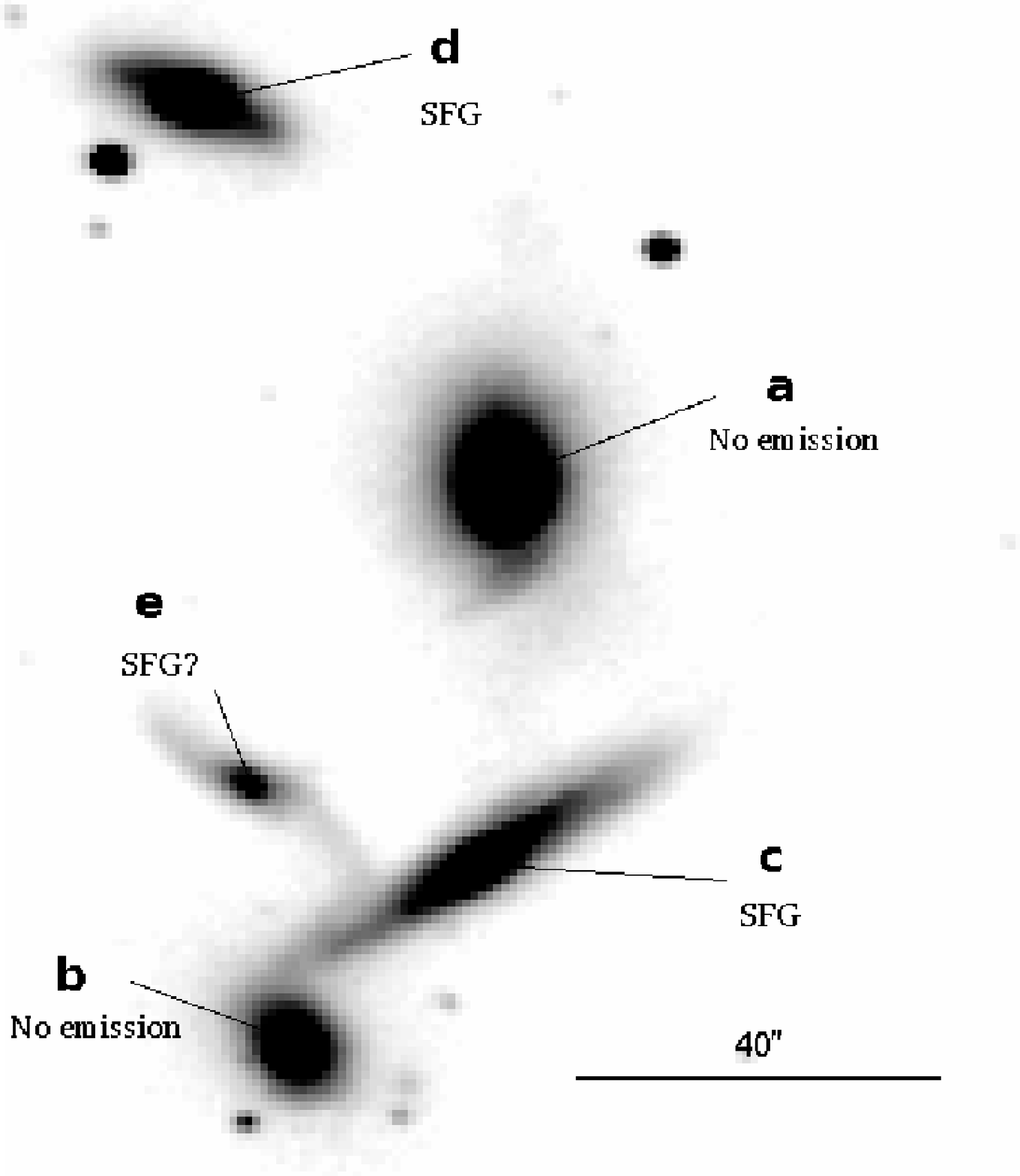}
\caption{Same as Figure~2 for HCG 40.
The fainter grey structures shown in this image have a mean S/N of 25.
\label{fig4}}
\end{figure}
\clearpage

\begin{figure}
\plotone{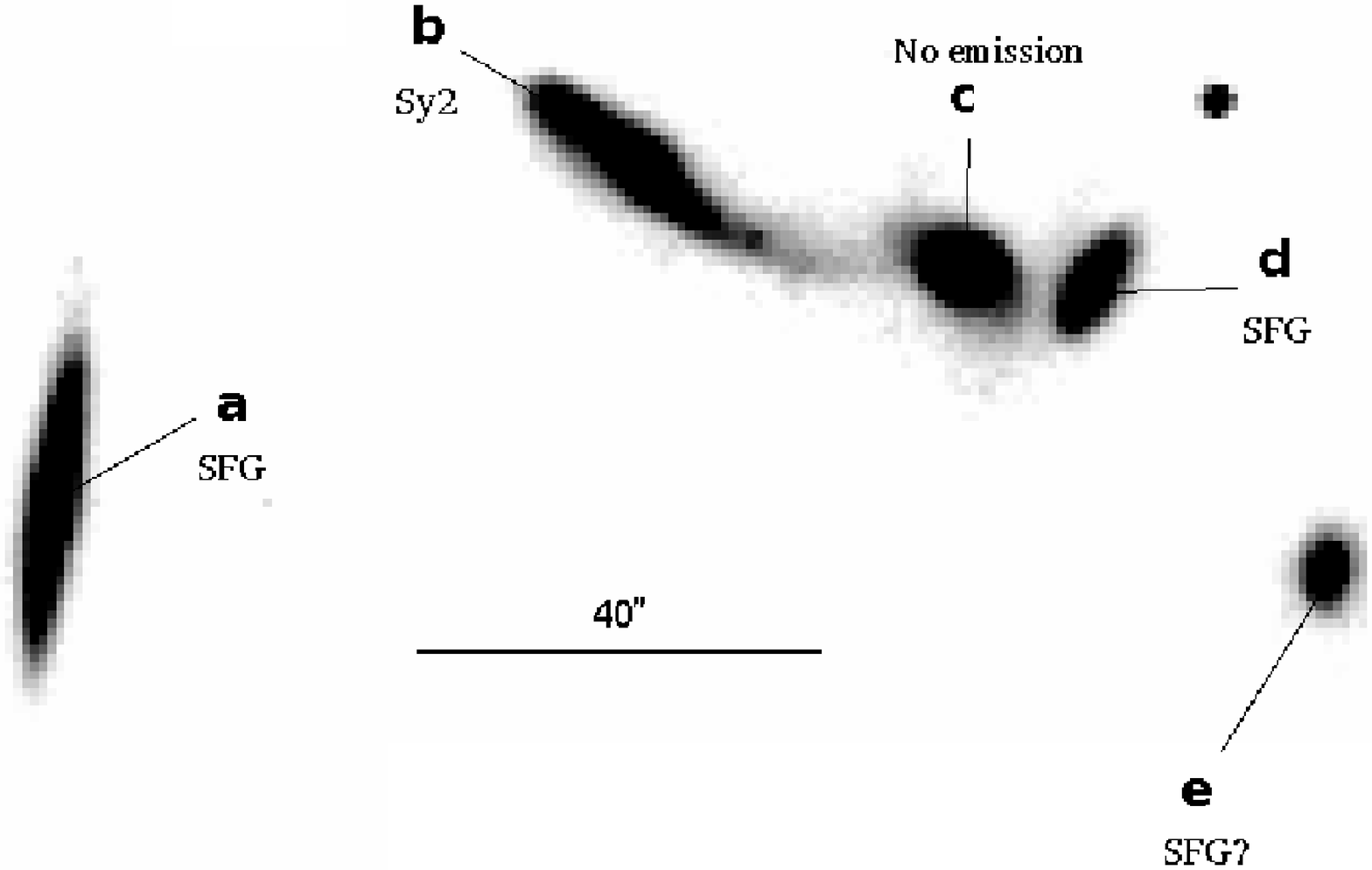}
\caption{Same as Figure~2 for HCG 56.
The fainter grey structures shown in this image have a mean S/N of 4.
\label{fig5}}
\end{figure}
\clearpage

\begin{figure}
\plotone{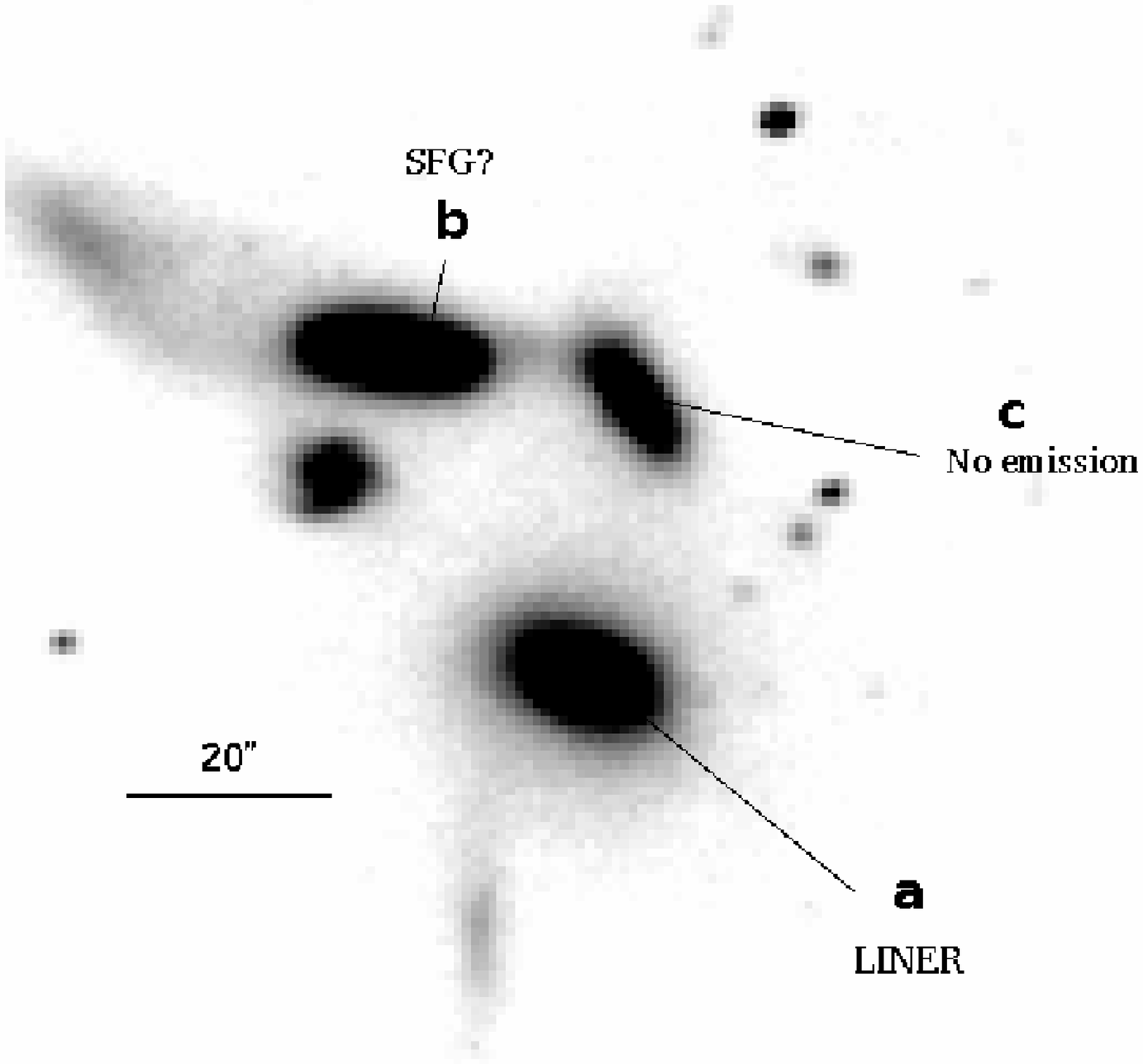}
\caption{Same as Figure~2 for HCG 79.
The fainter grey structures shown in this image have a mean S/N of 12.
\label{fig6}}
\end{figure}
\clearpage

\begin{figure}
\plotone{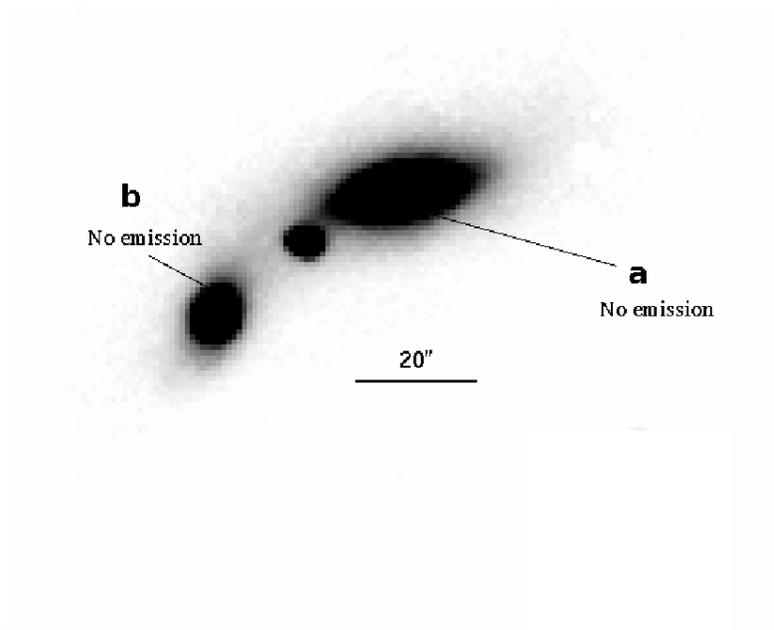}
\caption{Same as Figure~2 for HCG~98. The non-identified objects in the center is a foreground star.
The fainter grey structures shown in this image have a mean S/N of 20.
\label{fig7}}
\end{figure}
\clearpage

\begin{figure}
\plotone{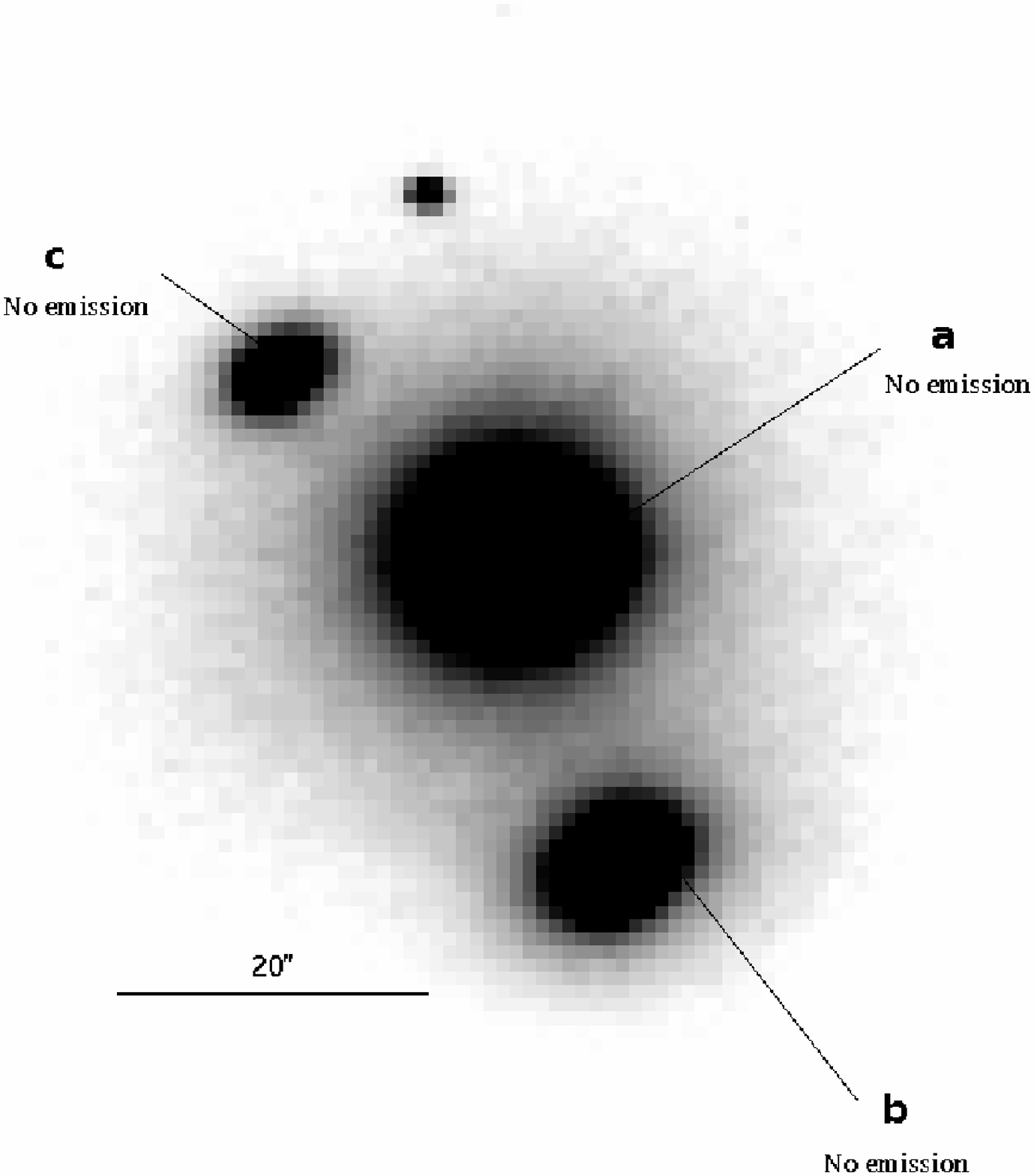}
\caption{Same as Figure~2 for HCG 74.
The fainter grey structures shown in this image have a mean S/N of 27.
\label{fig8}}
\end{figure}
\clearpage

\begin{figure}
\plotone{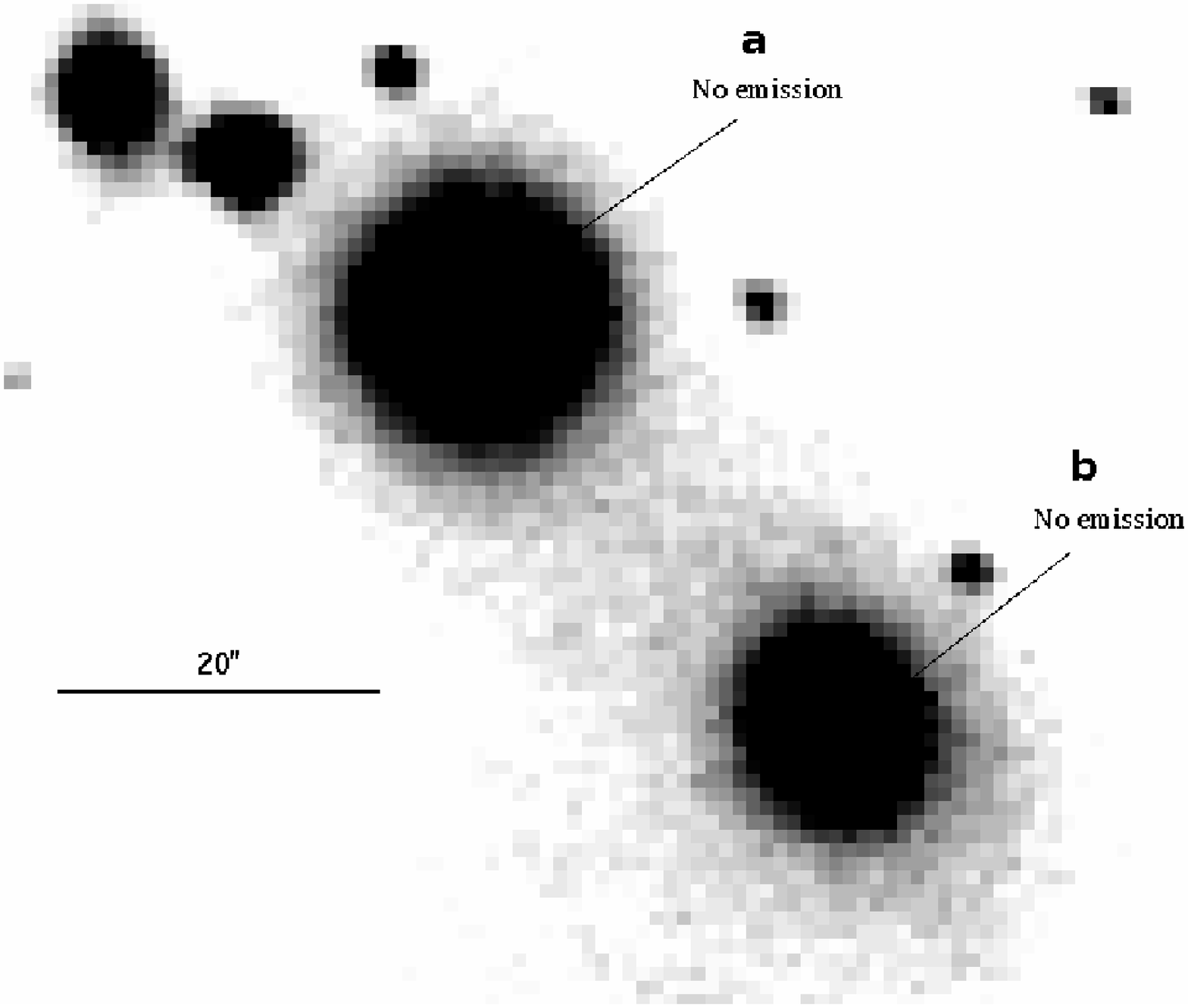}
\caption{Same as Figure~2 for HCG 94.
The fainter grey structures shown in this image have a mean S/N of 2.
\label{fig9}}
\end{figure}

\clearpage

\begin{figure}
\epsscale{.70} 
\plotone{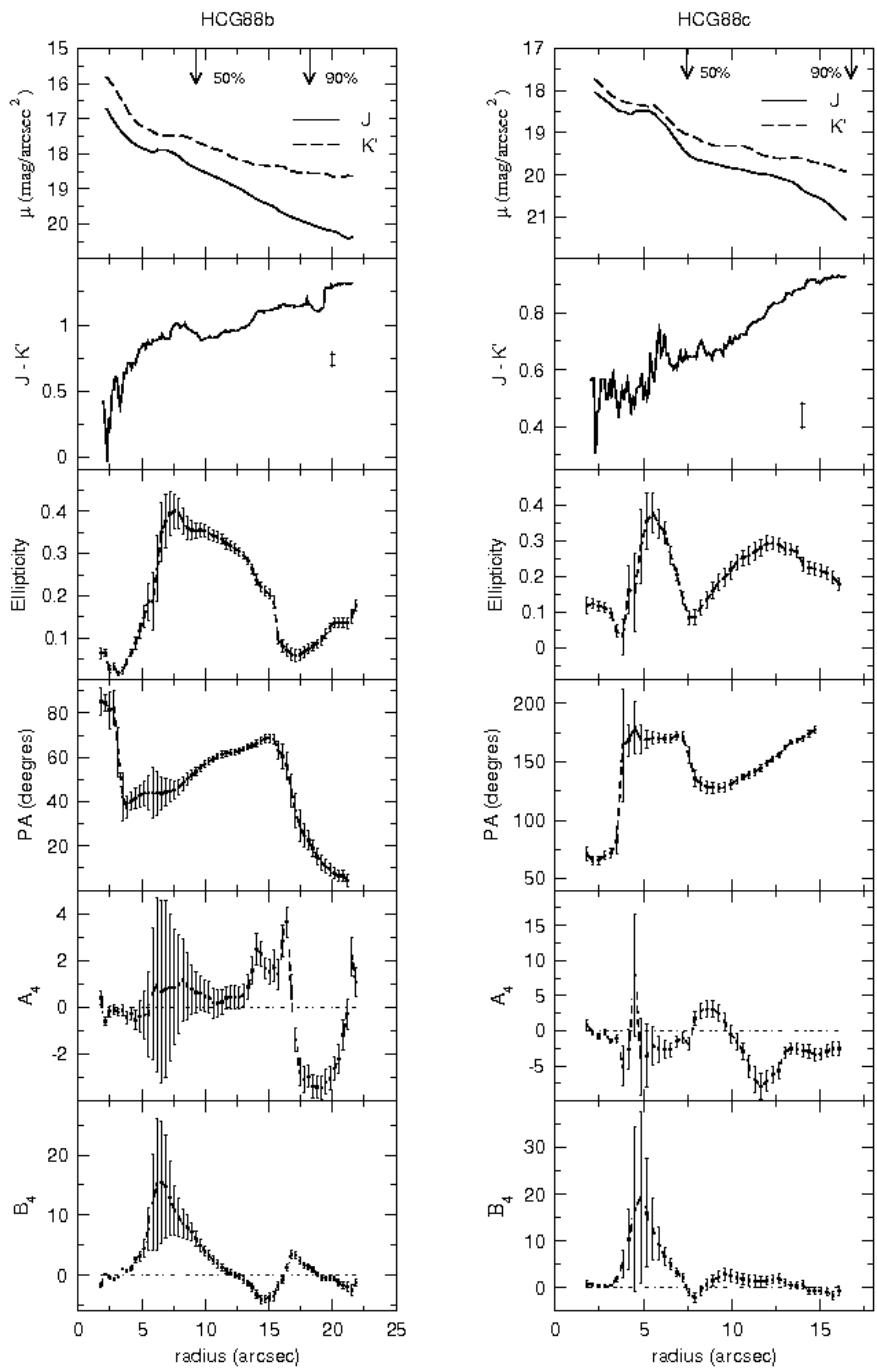} \caption{Results of the fitted
ellipses analysis. The graphics show, from top to bottom, the
surface brightness profiles (J and K'), the color profile (with
typical error bar) and variations with radius of the ellipticity,
position angle, and the fourth order Fourier coefficients. The two
arrows in the upper graph indicate the radii containing 50\% and
90\% of the light.\label{fig10}}
\end{figure}
\clearpage

\begin{figure}
\epsscale{.70}
\plotone{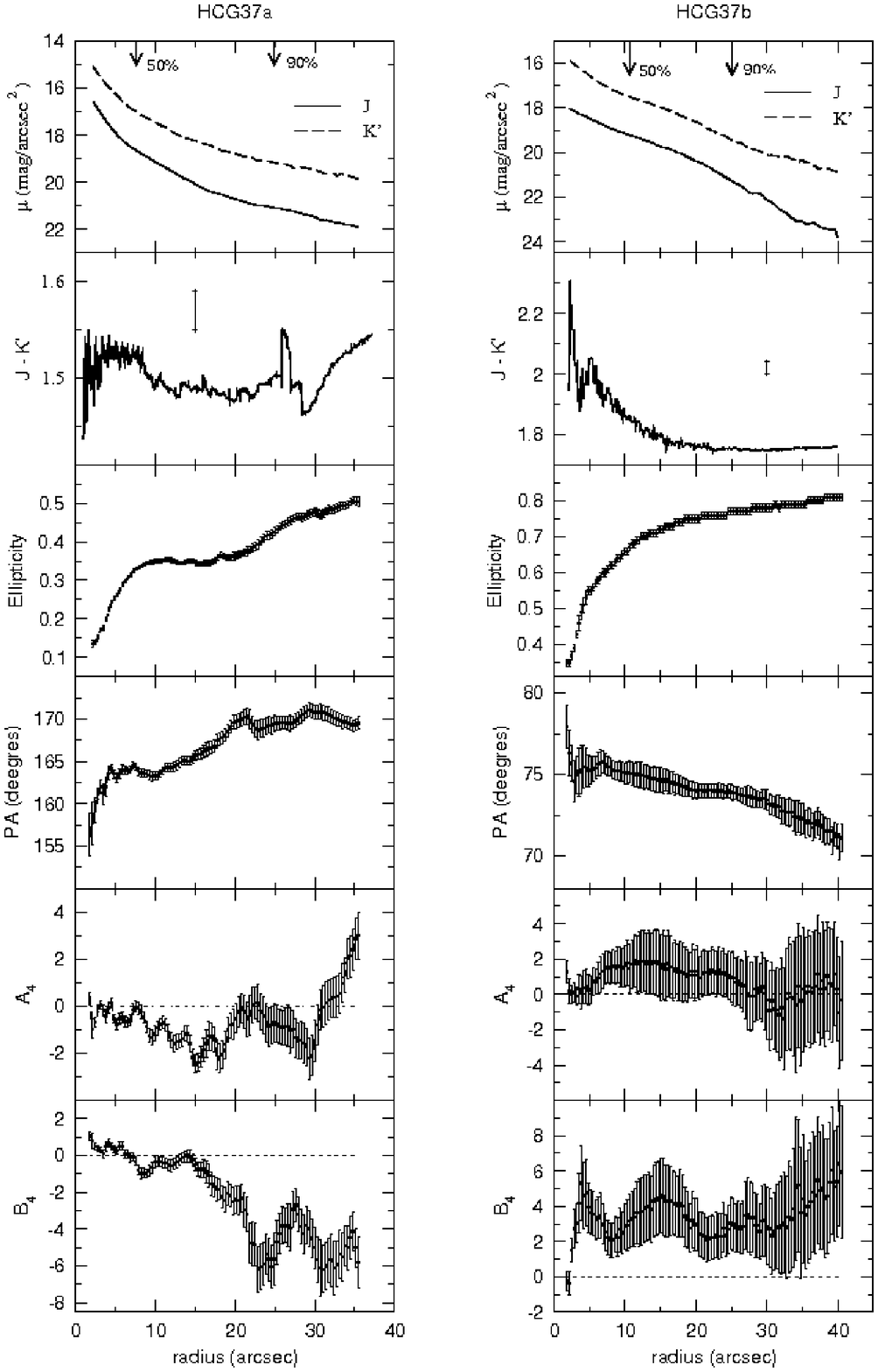}
\caption{Same as figure 10.\label{fig11}}
\end{figure}
\clearpage

\begin{figure}
\plotone{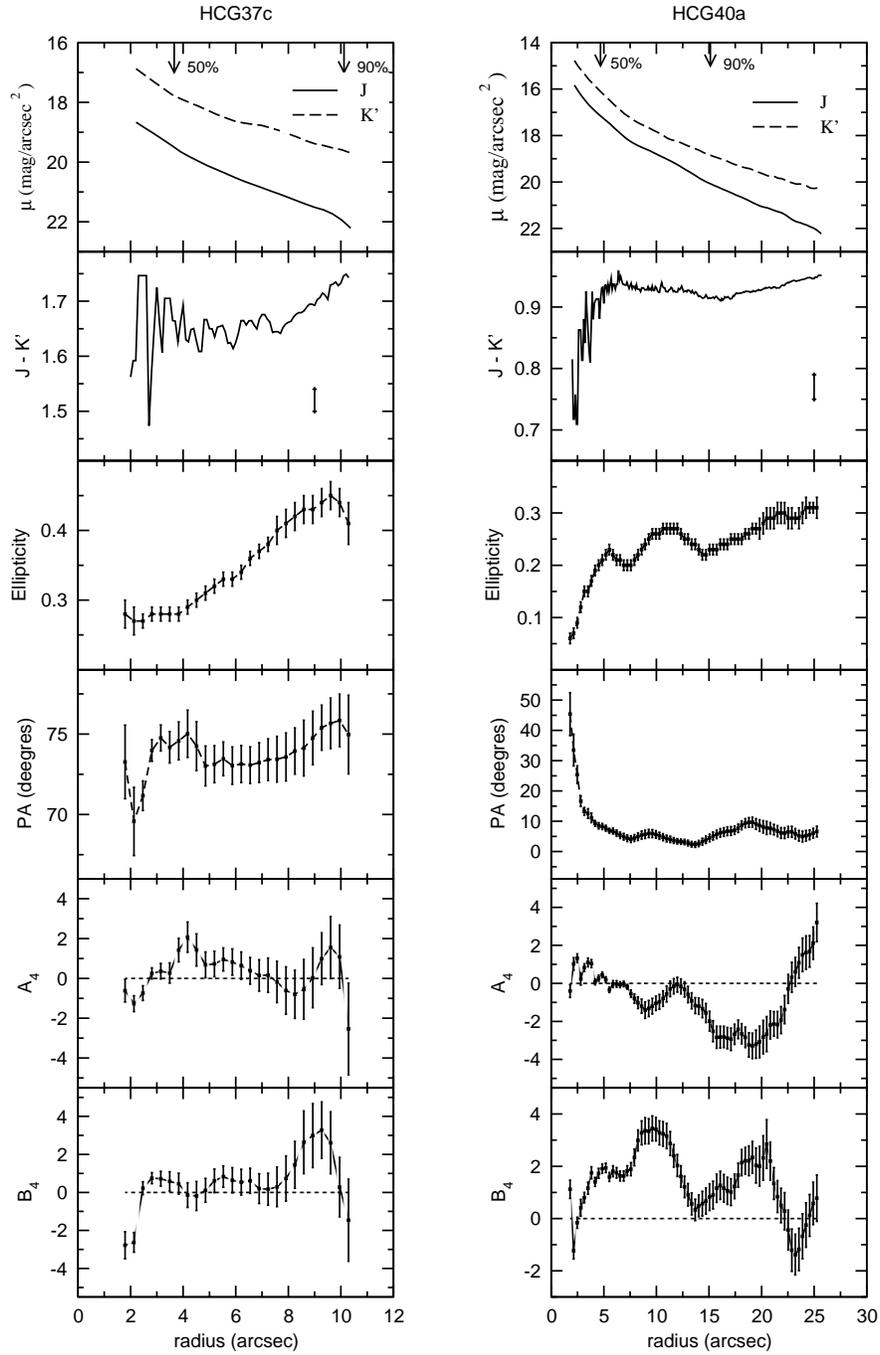}
\caption{Same as figure 10.\label{fig12}}
\end{figure}
\clearpage

\begin{figure}
\plotone{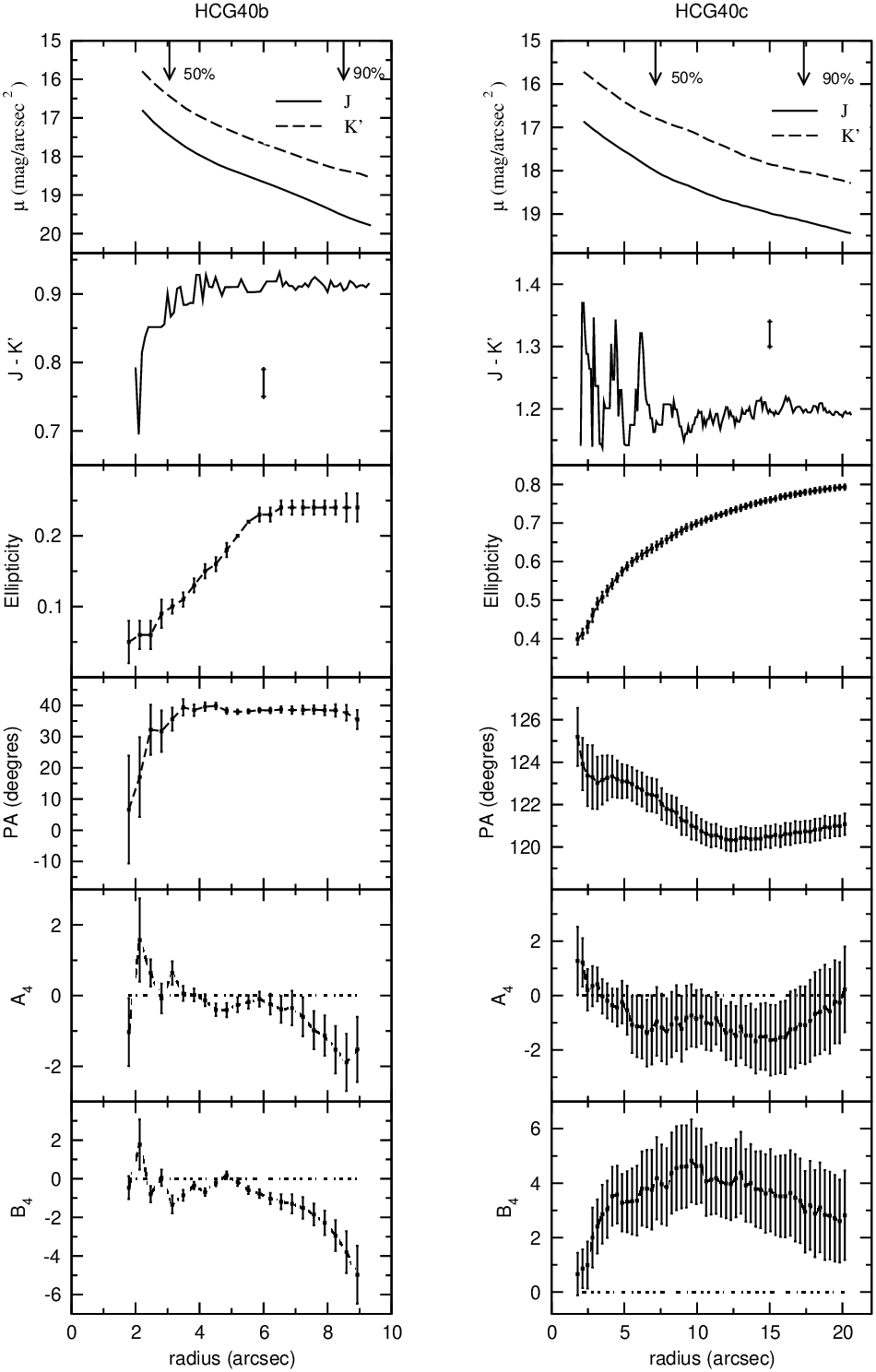}
\caption{Same as figure 10.\label{fig13}}
\end{figure}
\clearpage

\begin{figure}
\plotone{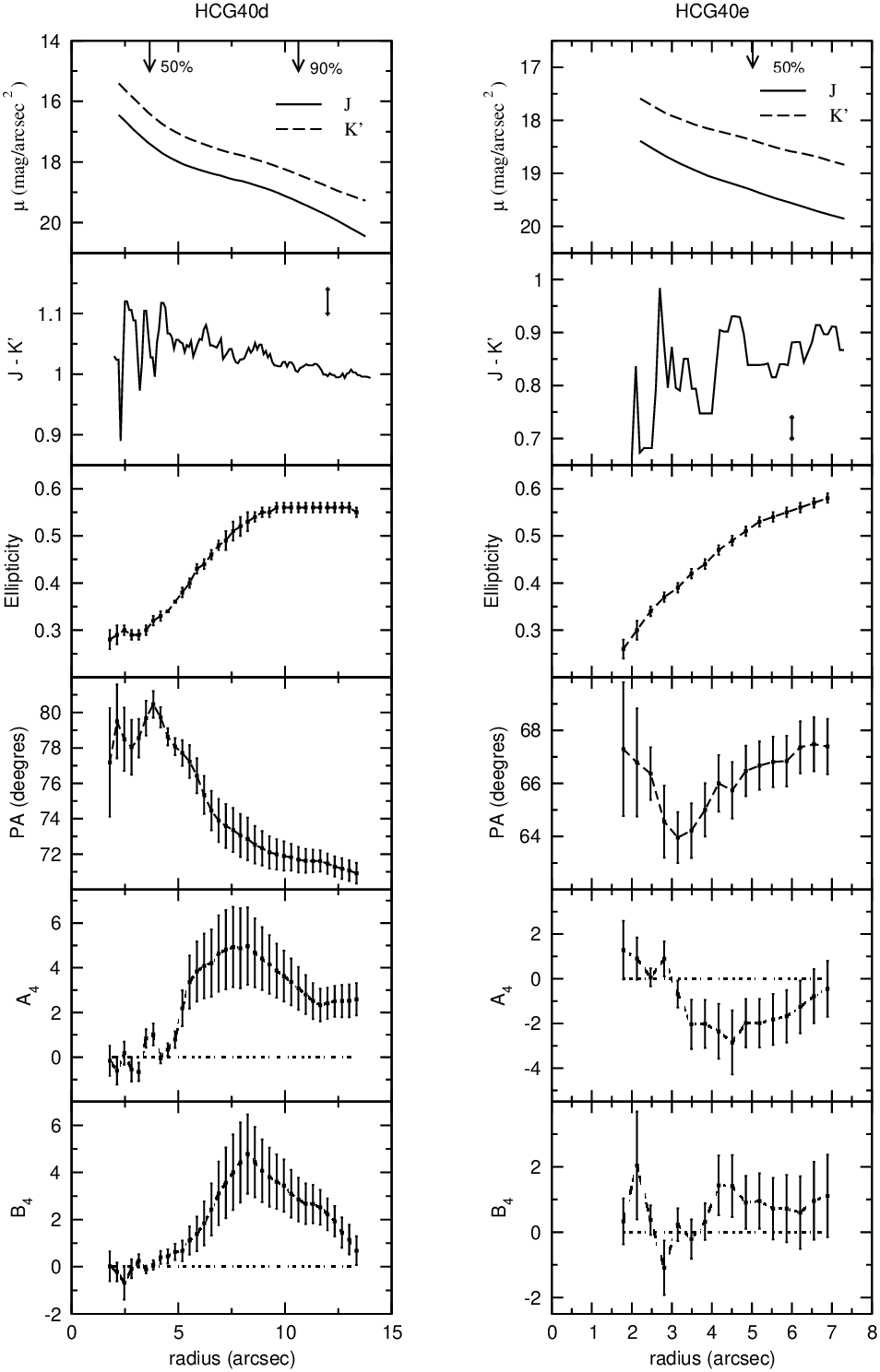}
\caption{Same as figure 10.\label{fig14}}
\end{figure}
\clearpage

\begin{figure}
\plotone{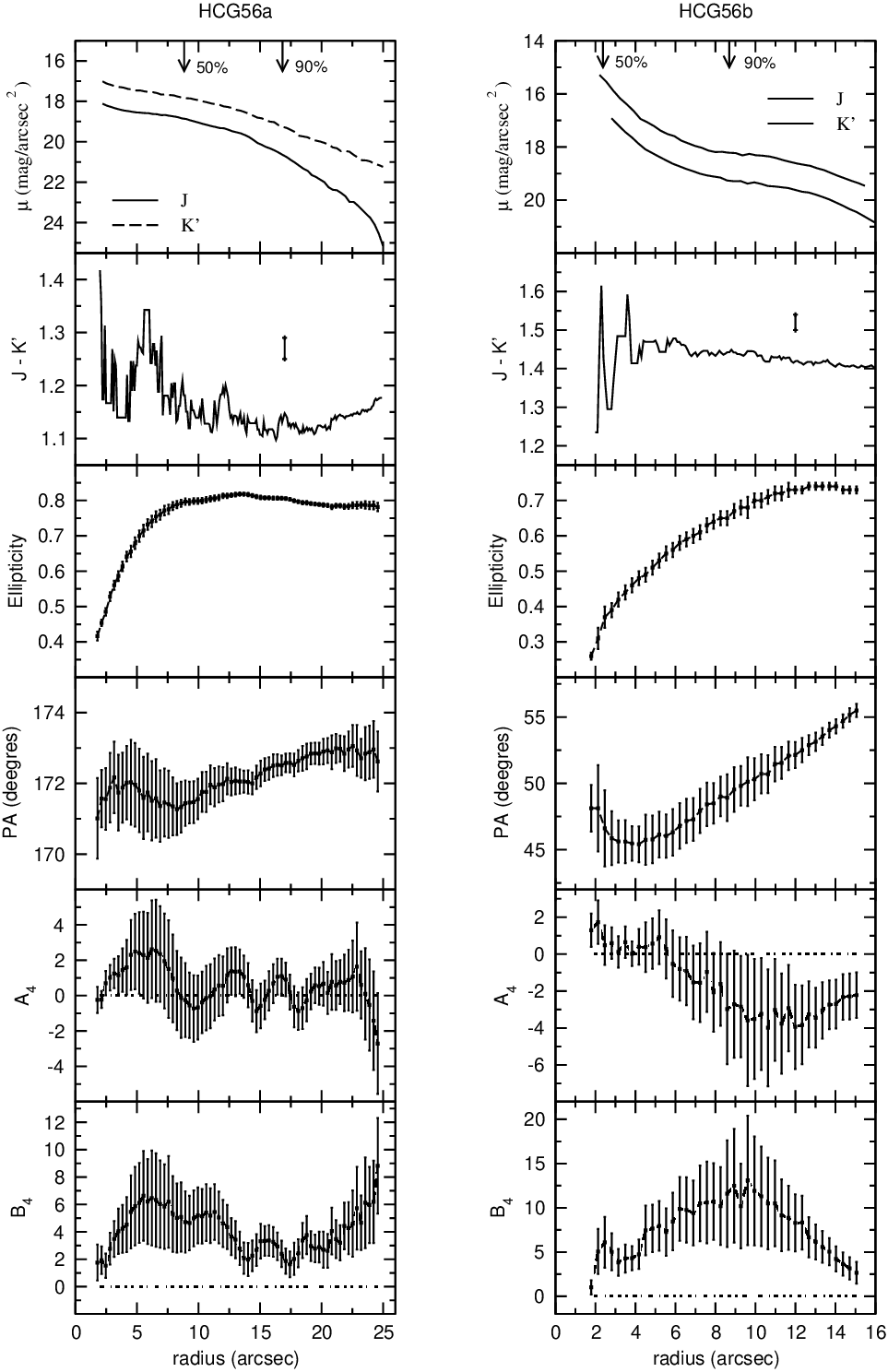}
\caption{Same as figure 10.\label{fig15}}
\end{figure}
\clearpage

\begin{figure}
\plotone{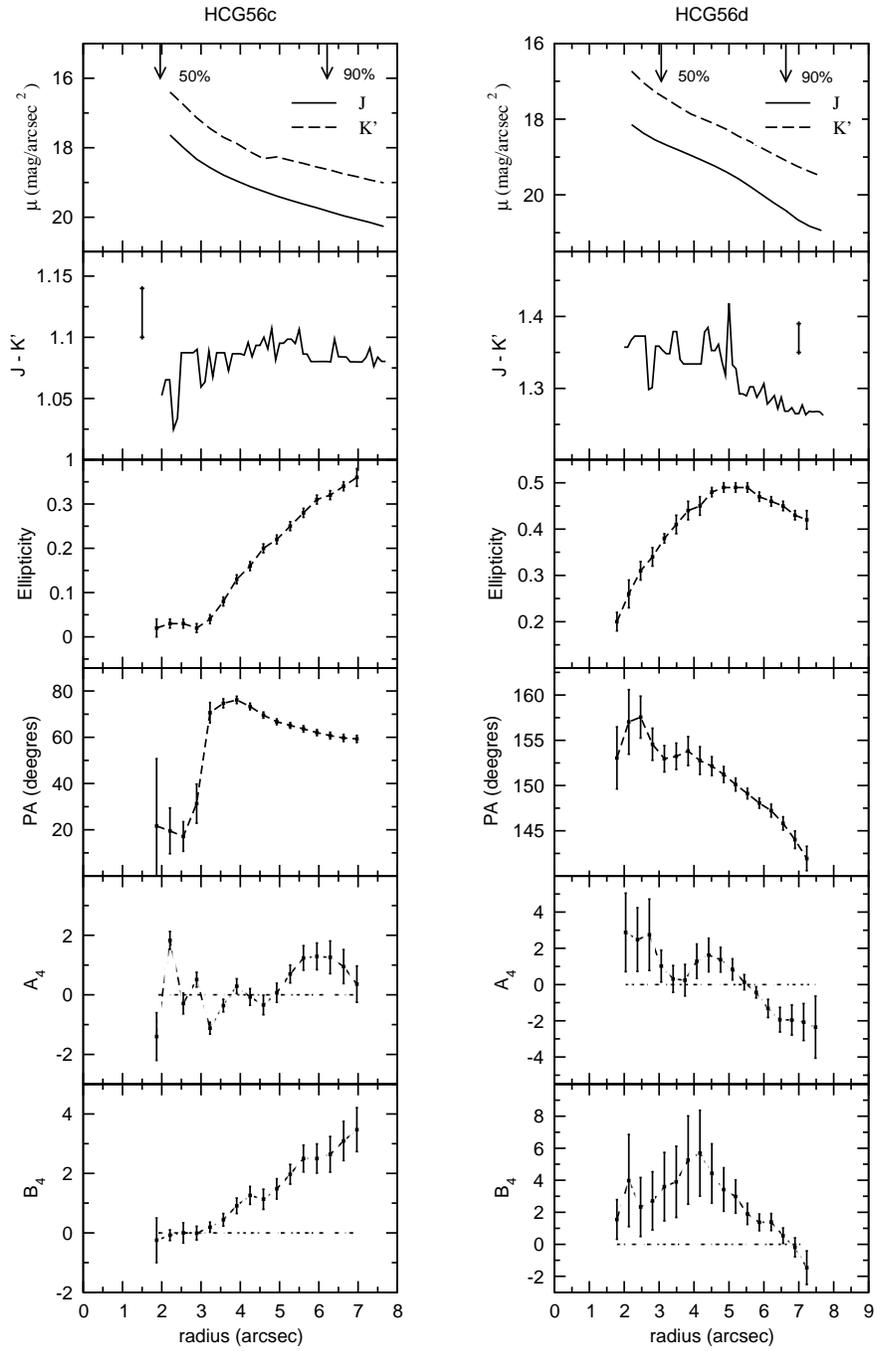}
\caption{Same as figure 10.\label{fig16}}
\end{figure}
\clearpage

\begin{figure}
\plotone{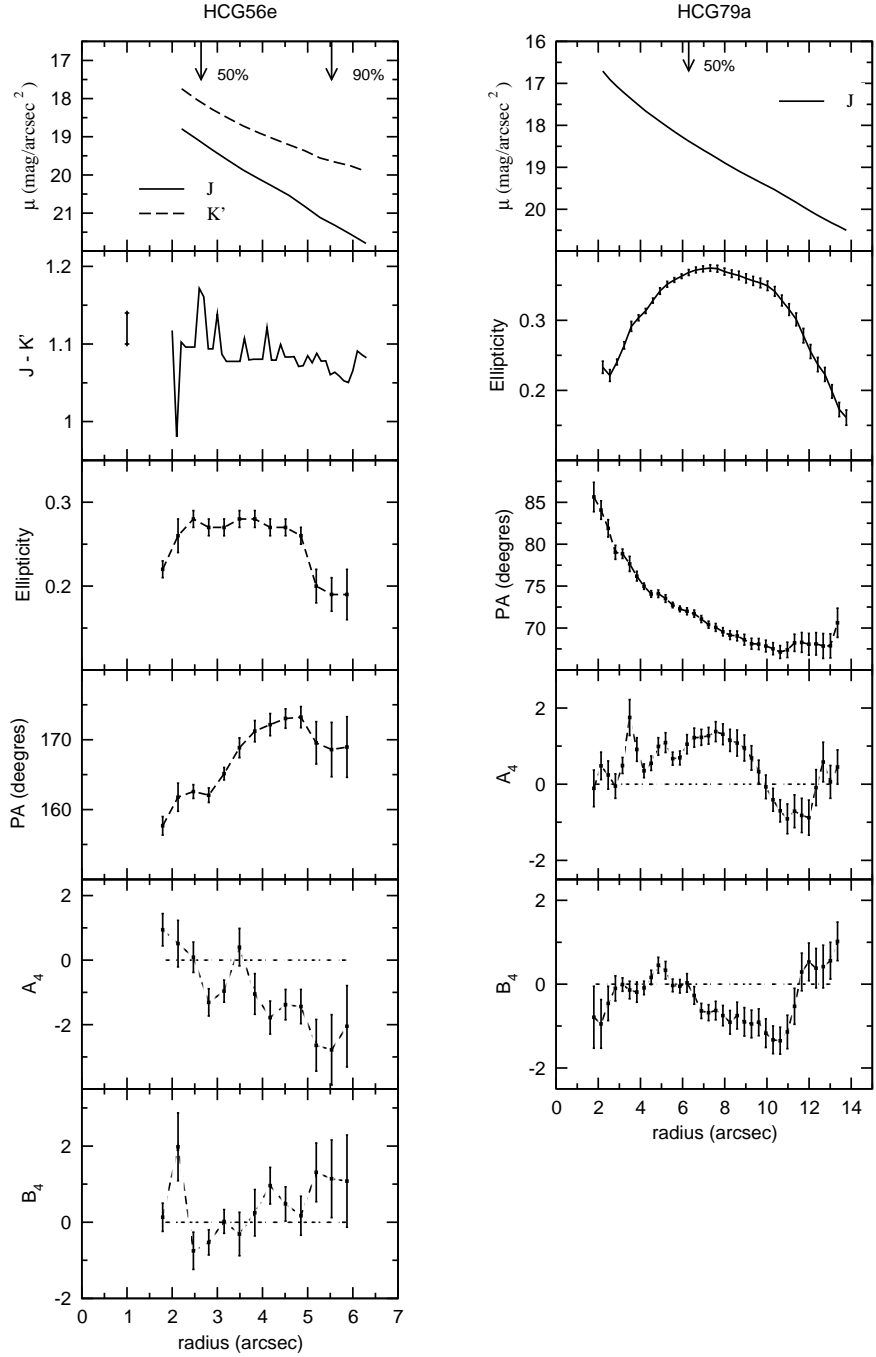} \caption{Same as figure 10 for HCG~56e. The K'
image is missing in the case of HCG~79a.\label{fig17}}
\end{figure}
\clearpage

\begin{figure}
\plotone{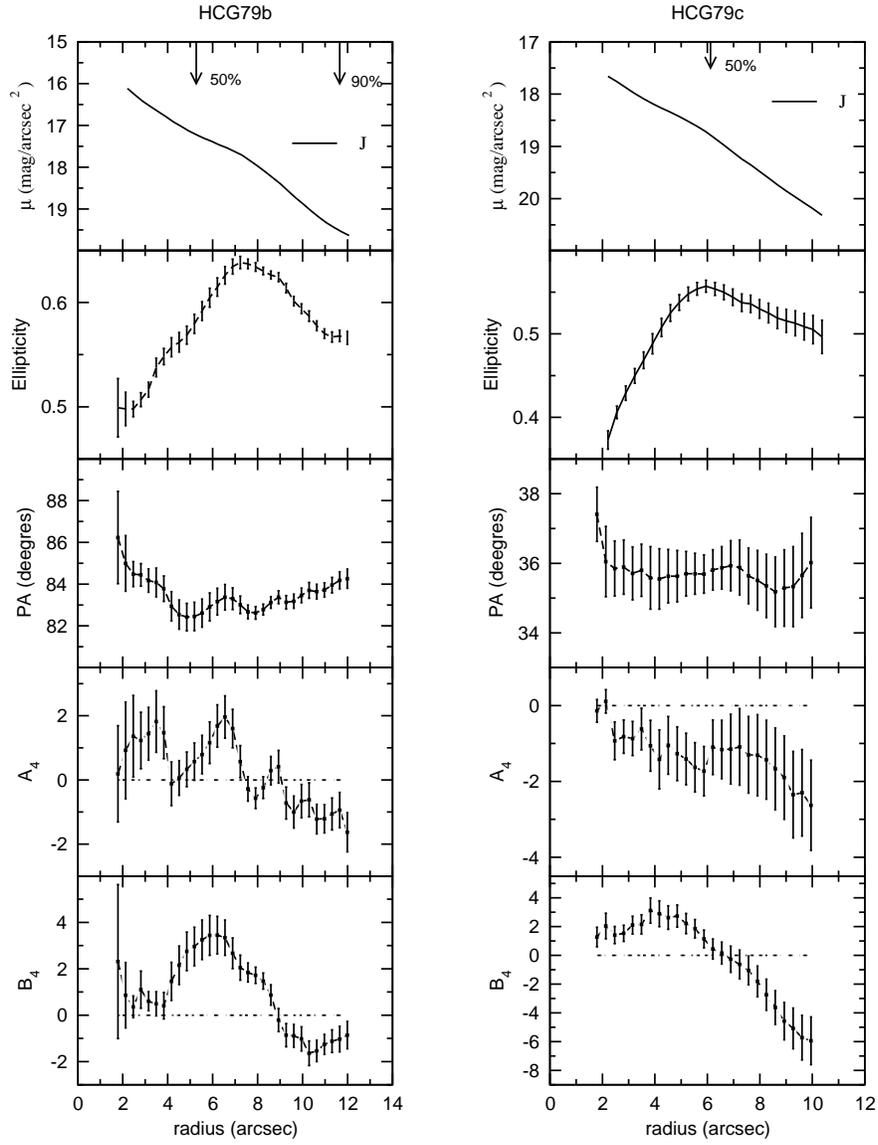}
\caption{Same as figure 10 , except that the K' image is missing.\label{fig18}}
\end{figure}
\clearpage

\begin{figure}
\plotone{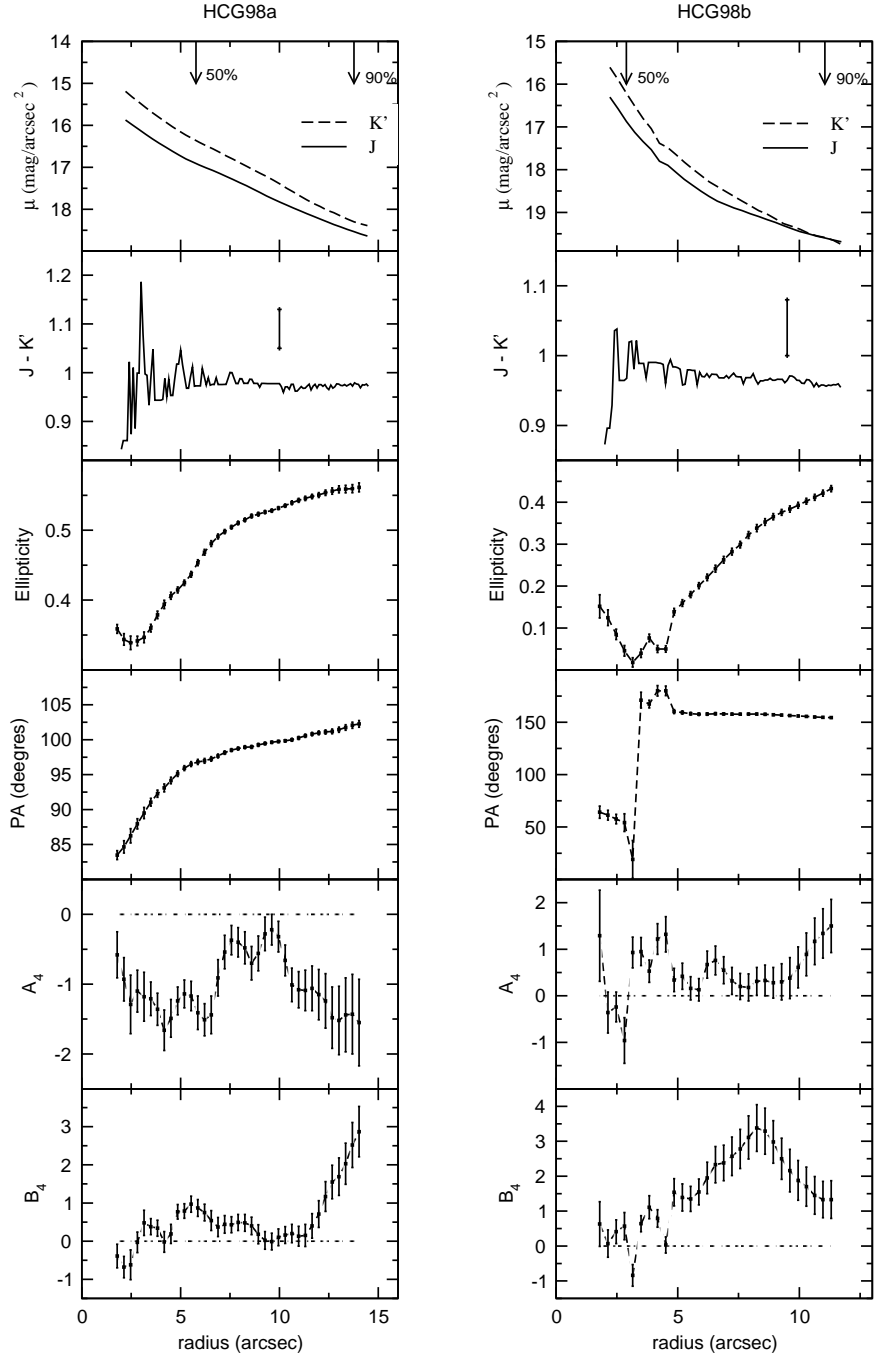}
\caption{Same as figure 10.\label{fig19}}
\end{figure}
\clearpage

\begin{figure}
\plotone{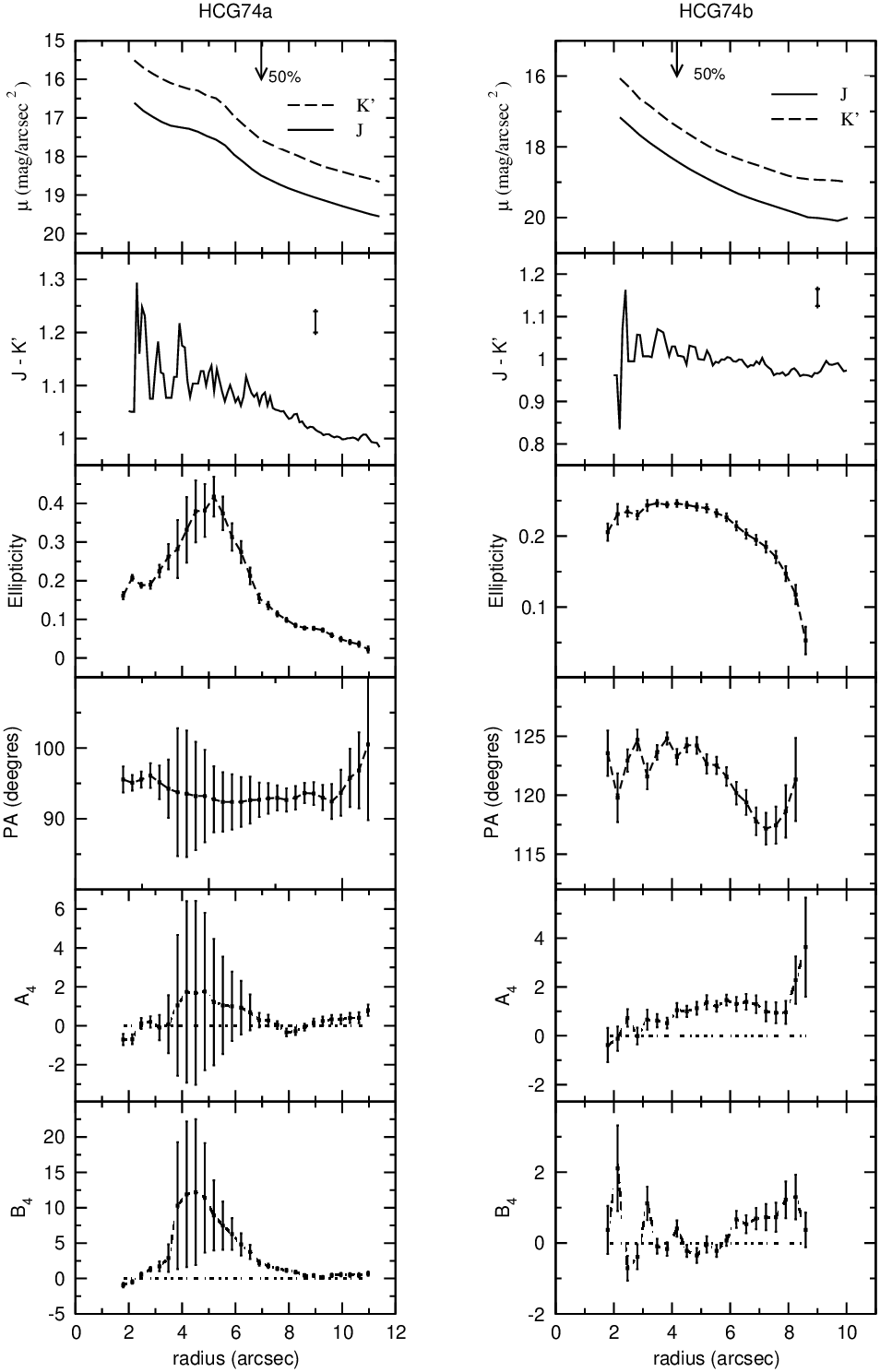}
\caption{Same as figure 10.\label{fig20}}
\end{figure}
\clearpage

\begin{figure}
\plotone{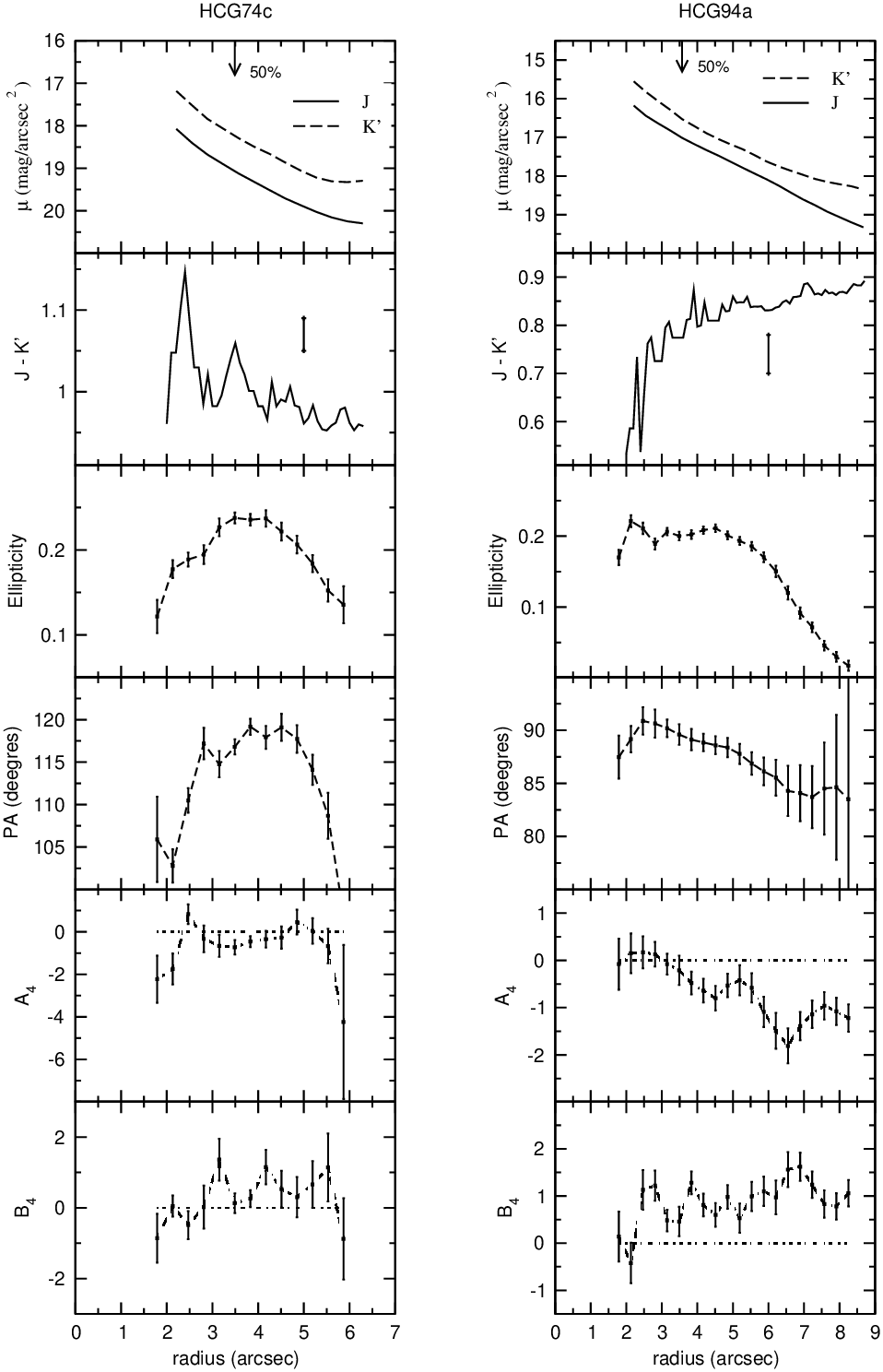}
\caption{Same as figure 10.\label{fig21}}
\end{figure}
\clearpage

\begin{figure}
\epsscale{.35}
\plotone{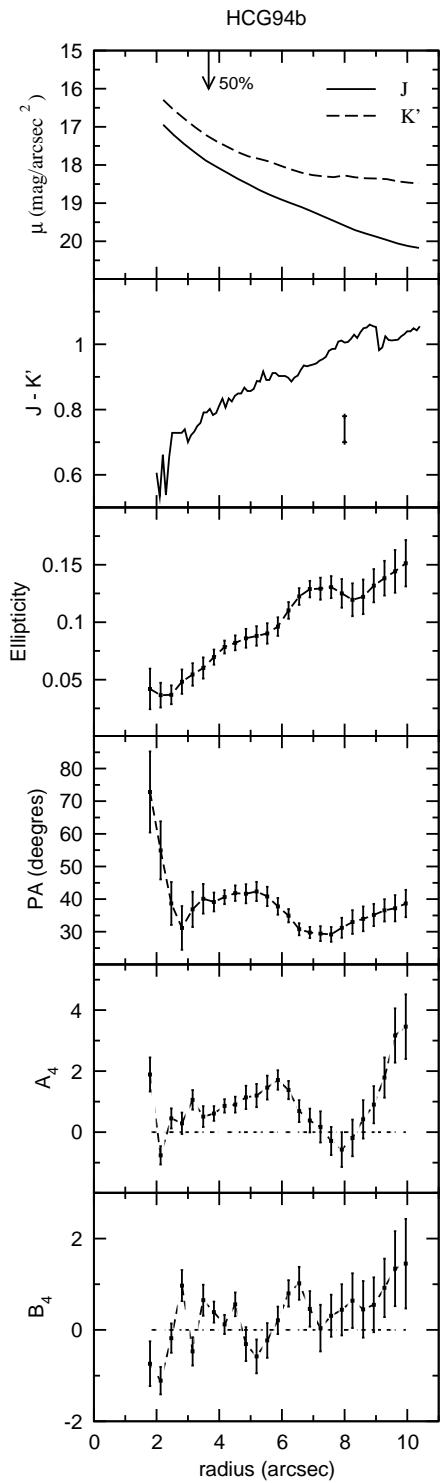}
\caption{Same as figure 10.\label{fig22}}
\end{figure}

\clearpage

\begin{figure}
\epsscale{0.9}
\plotone{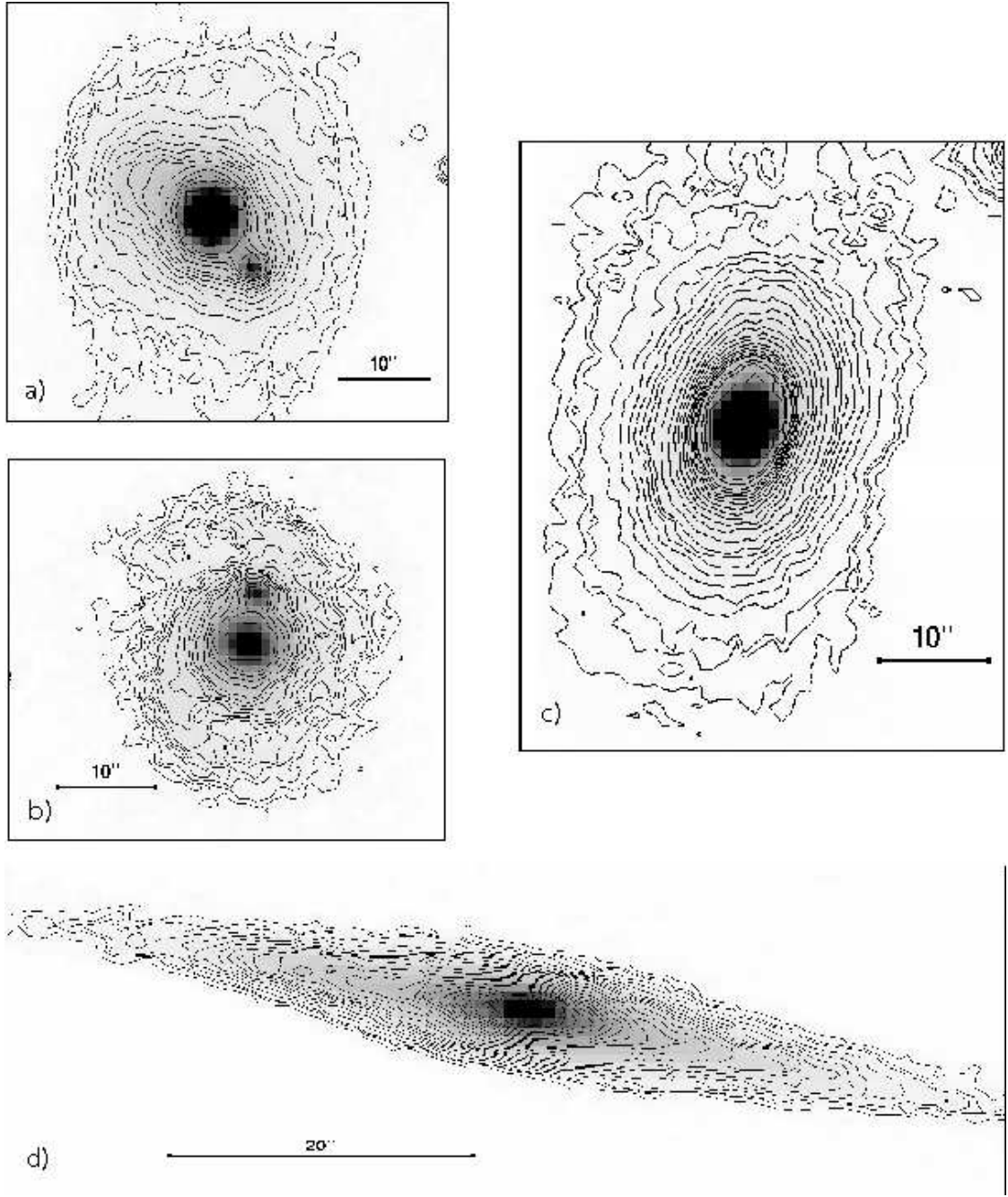}
\caption{Photometrically uncalibrated isophotes (units in ADU), overlayed on the negative J images
of the galaxies in our sample. The levels decrease with radius by ~ 5\% (compared to the
inner isophote). As usual, north is up and east to the left.
The galaxies are: a) HCG~88b, b) HCG~88c, c) HCG~37a, and d) HCG~37b.
\label{fig23}}
\end{figure}
\clearpage

\begin{figure}
\plotone{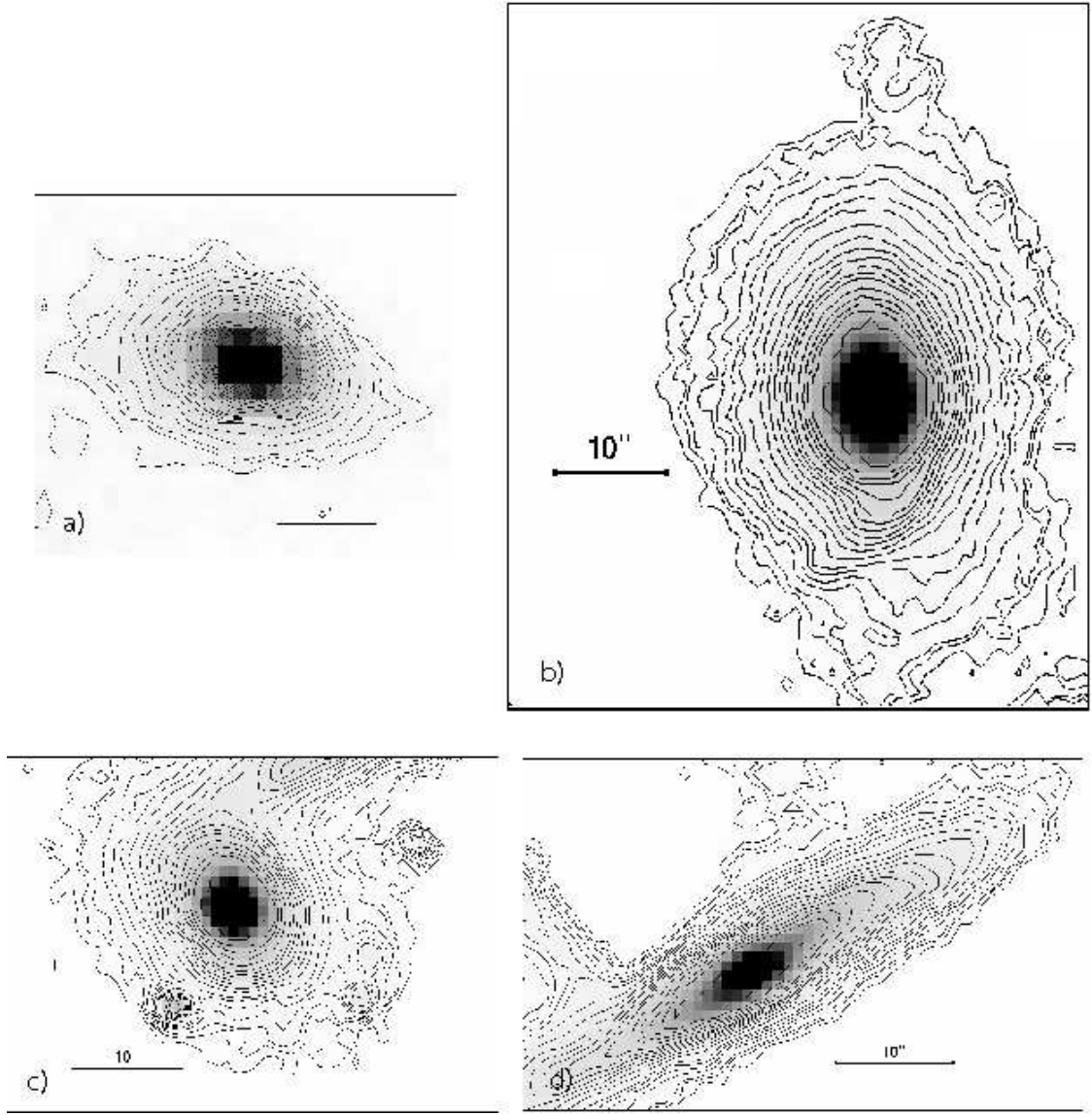}
\caption{Same as Figure~23 for a) HCG~37c, b)
HCG~40a, c) HCG~40b, and d) HCG~40c.\label{fig24}}
\end{figure}
\clearpage

\begin{figure}
\plotone{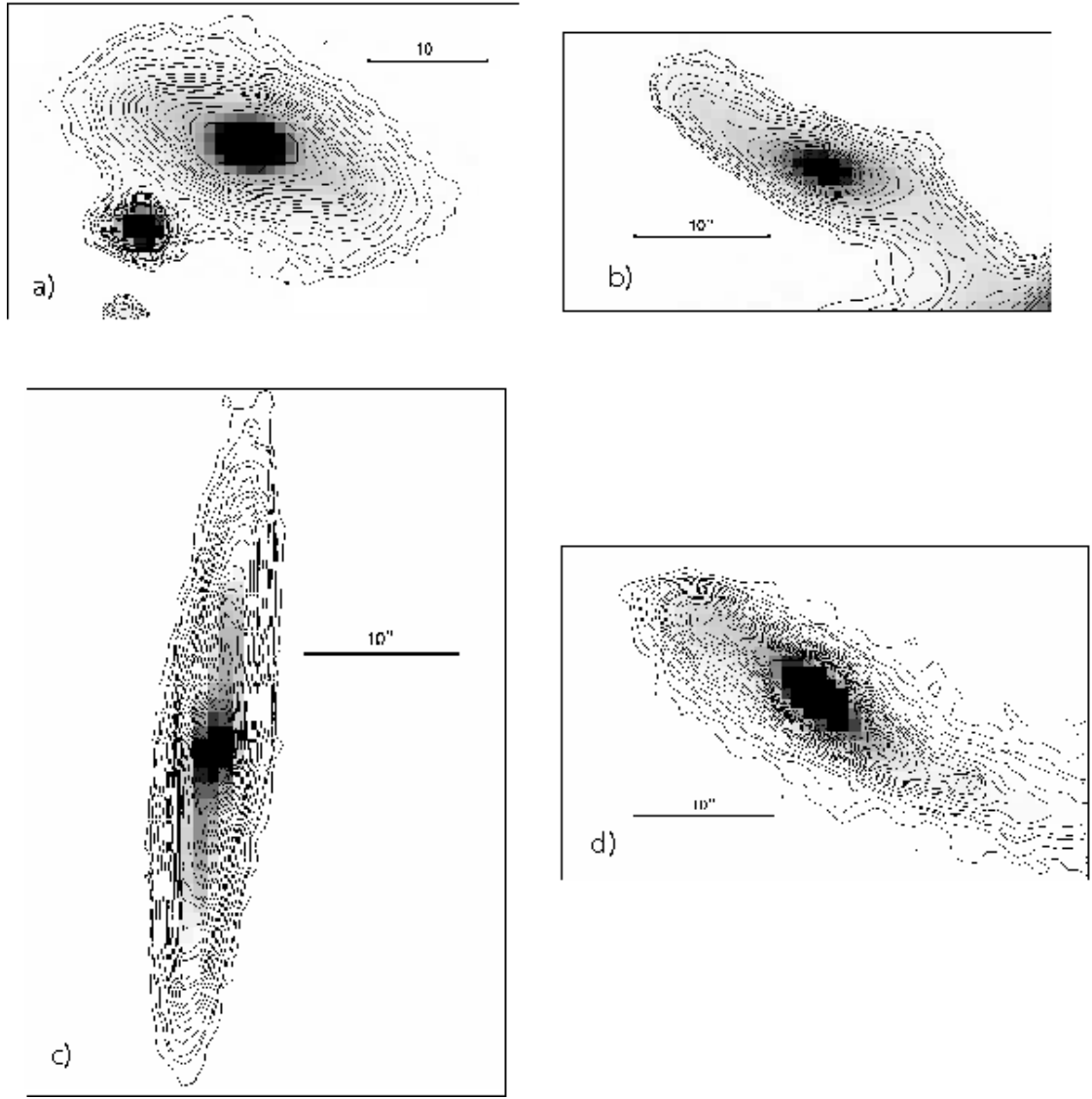}
\caption{Same as Figure~23 for a) HCG~40d, b)
HCG~40e, c) HCG~56a, and d) HCG~56b.\label{fig25}}
\end{figure}
\clearpage

\begin{figure}
\plotone{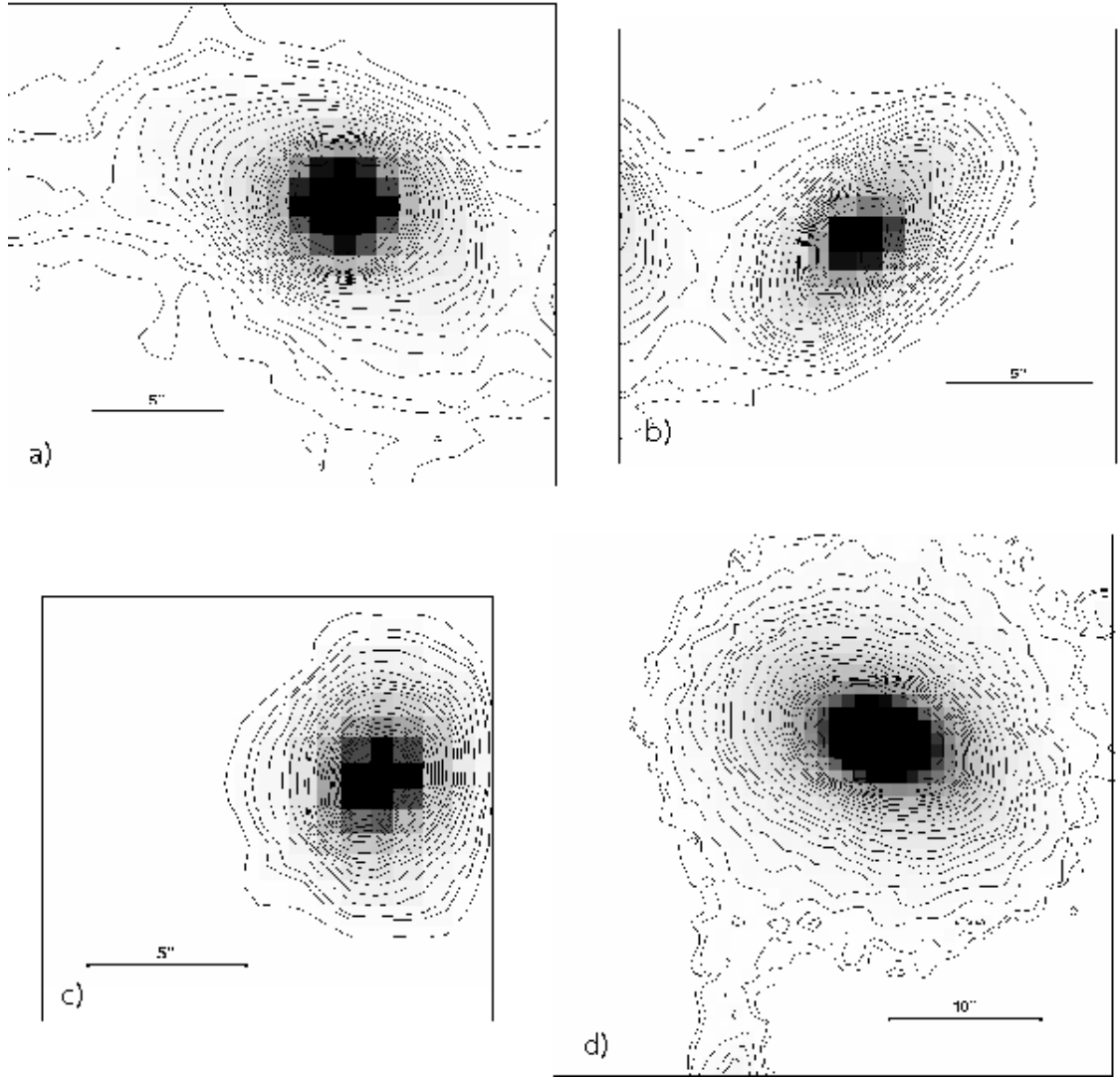}
\caption{Same as Figure~23 for a) HCG~56c, b)
HCG~56d, c) HCG~56e, and d) HCG~79a.\label{fig26}}
\end{figure}
\clearpage

\begin{figure}
\plotone{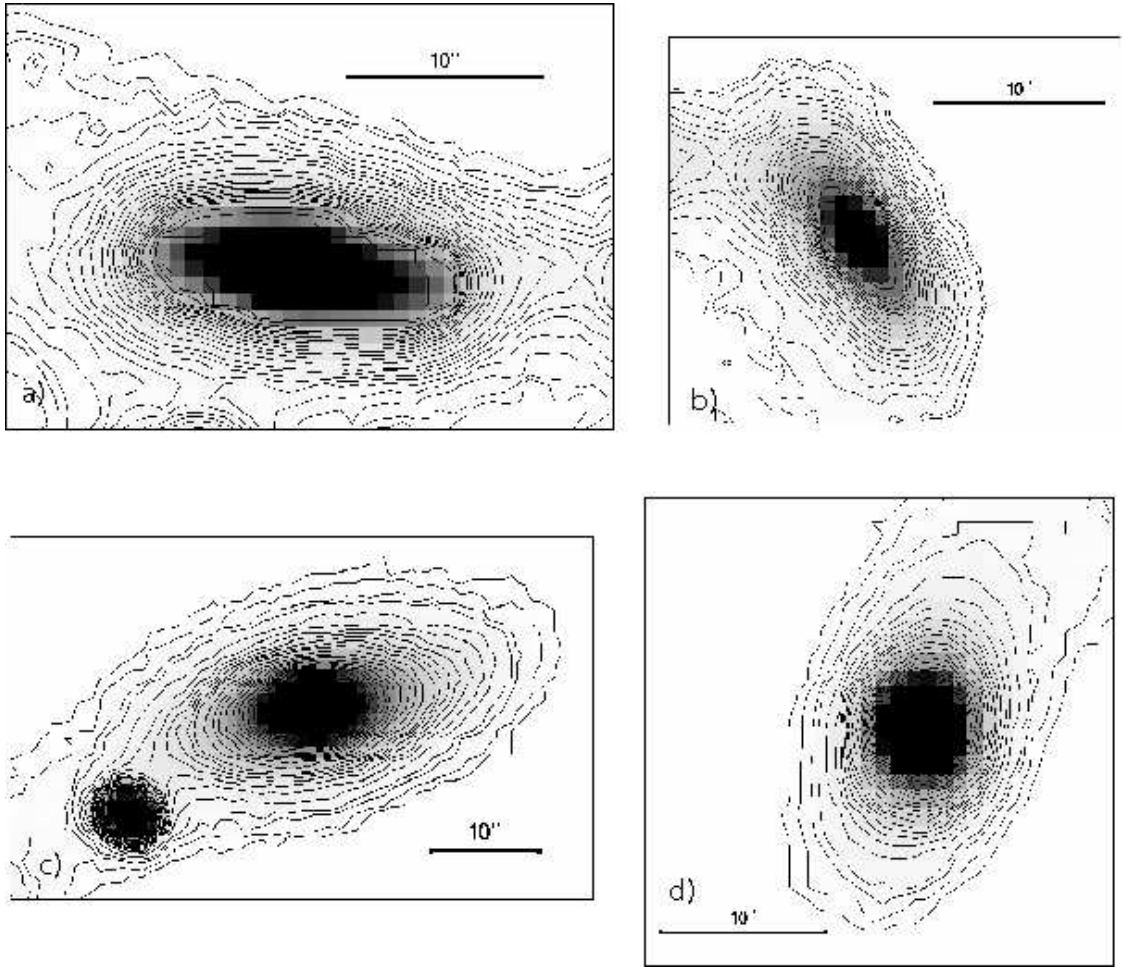}
\caption{Same as Figure~23 for a) HCG~79b, b)
HCG~79c, c) HCG~98a, and d) HCG~98b.\label{fig27}}
\end{figure}
\clearpage

\begin{figure}
\plotone{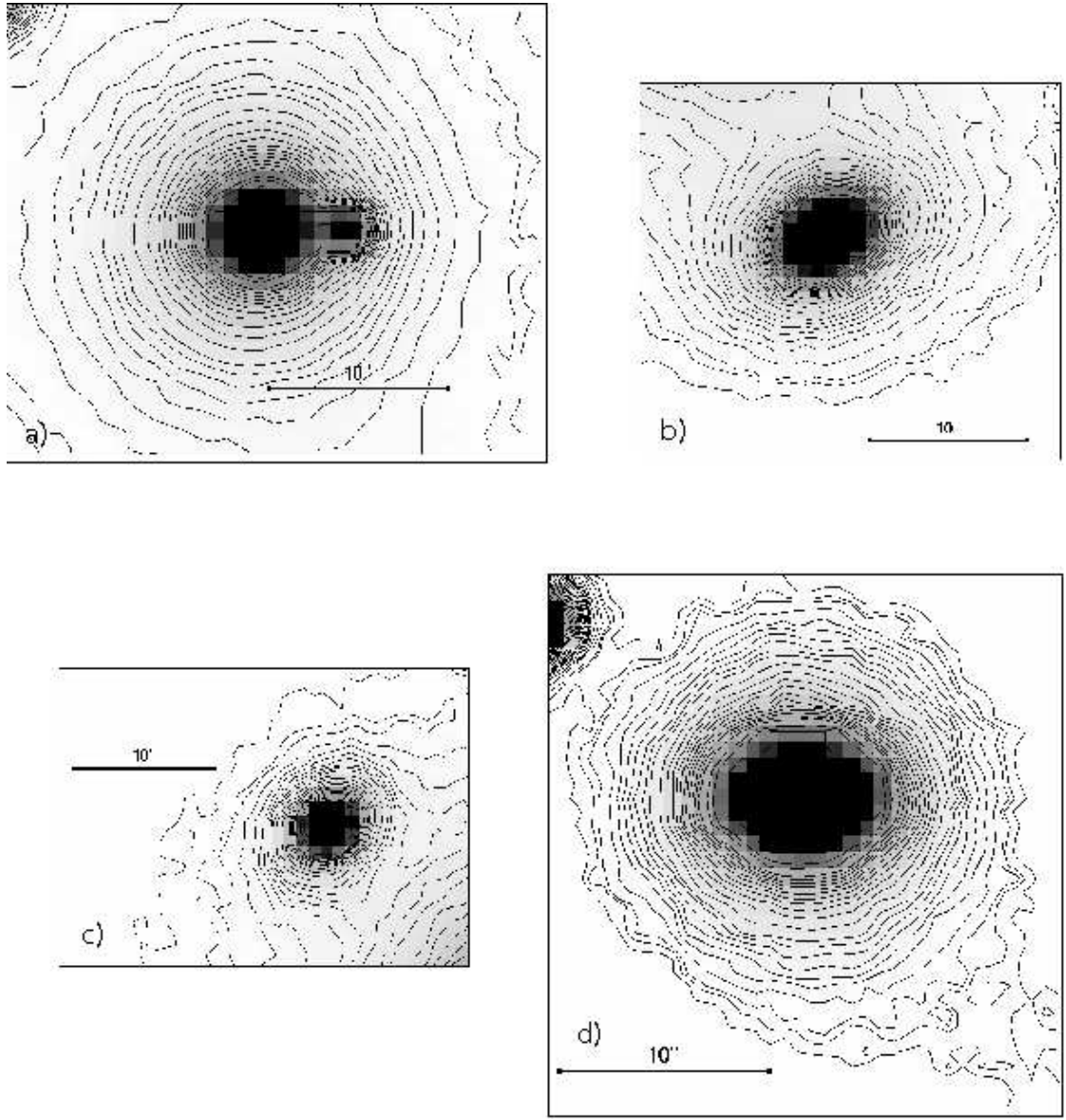}
\caption{Same as Figure~23 for a) HCG~74a, b)
HCG~74b, c) HCG~74c, and d) HCG~94a.\label{fig28}}
\end{figure}
\clearpage

\begin{figure}
\epsscale{.80}
\plotone{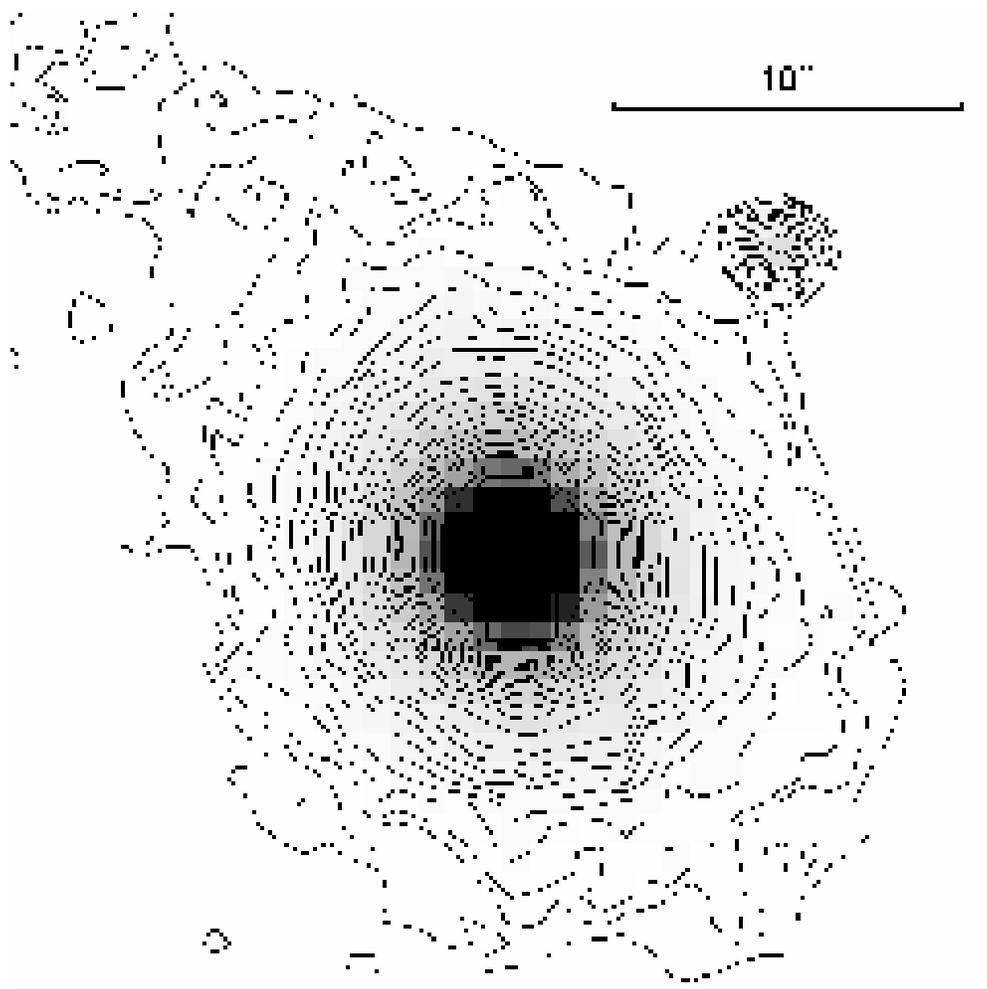}
\caption{Same as Figure~23 for
HCG~94b.\label{fig29}}
\end{figure}
\clearpage

\begin{figure}
\epsscale{0.8}
\plotone{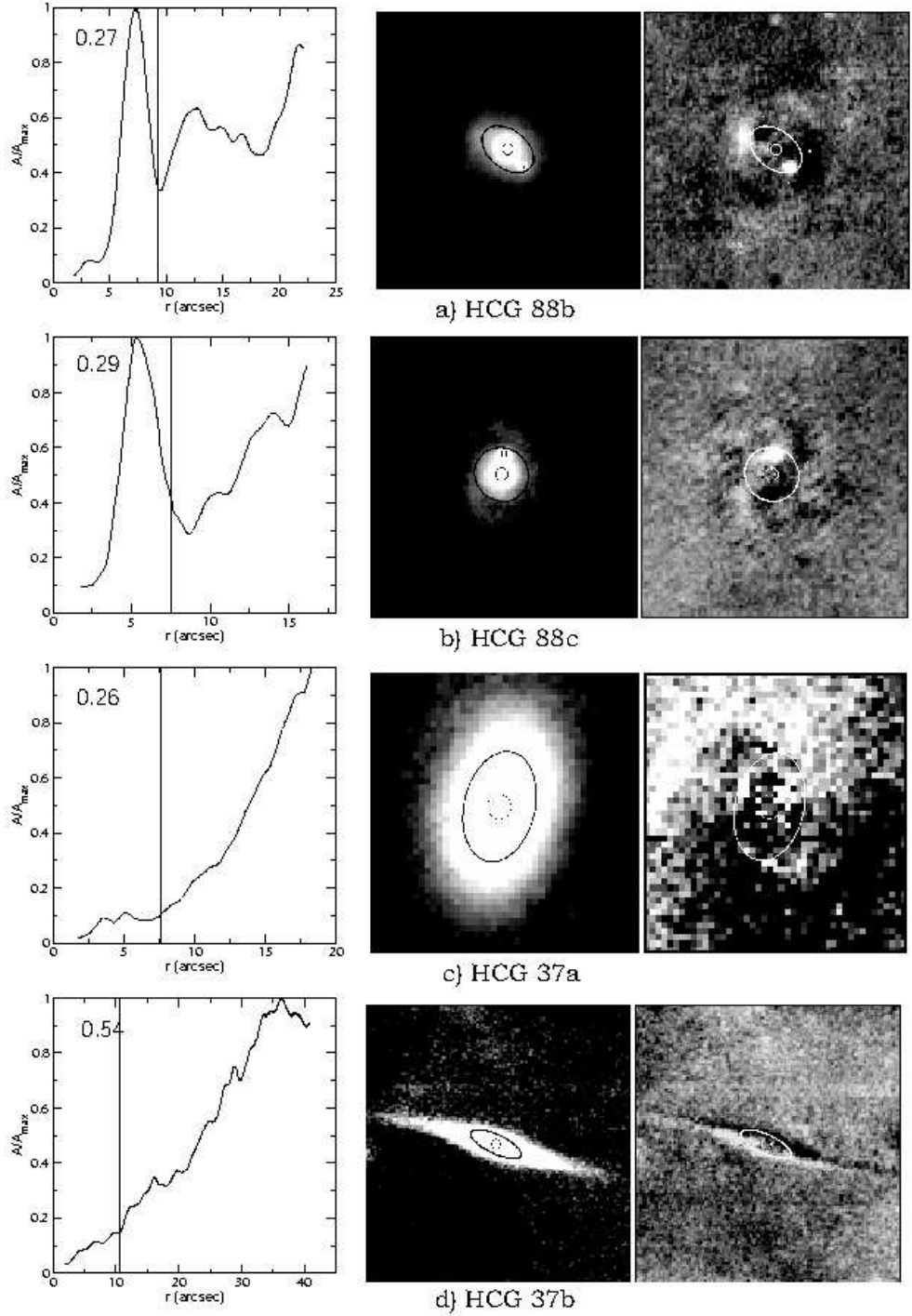}
\caption{Results of the asymmetry
analysis. The image in the center is the original image, cleaned (as
much as possible) of all foreground objects. The inner circle is the
size of the PSF and the ellipse is at isophote $r_{50}$. The image
to the right is the residual image obtained by subtracting the
rotated image from the original one. The two images are shown at the
same scale (north up and east to the left). The graphics to the left
give the relative asymmetry intensity ($A/A_{max}$) at different
radius. The maximum value is indicated in the upper left
corner. The vertical line is draw at $r_{50}$.\label{fig30}}
\end{figure}
\clearpage

\begin{figure}
\epsscale{0.8}
\plotone{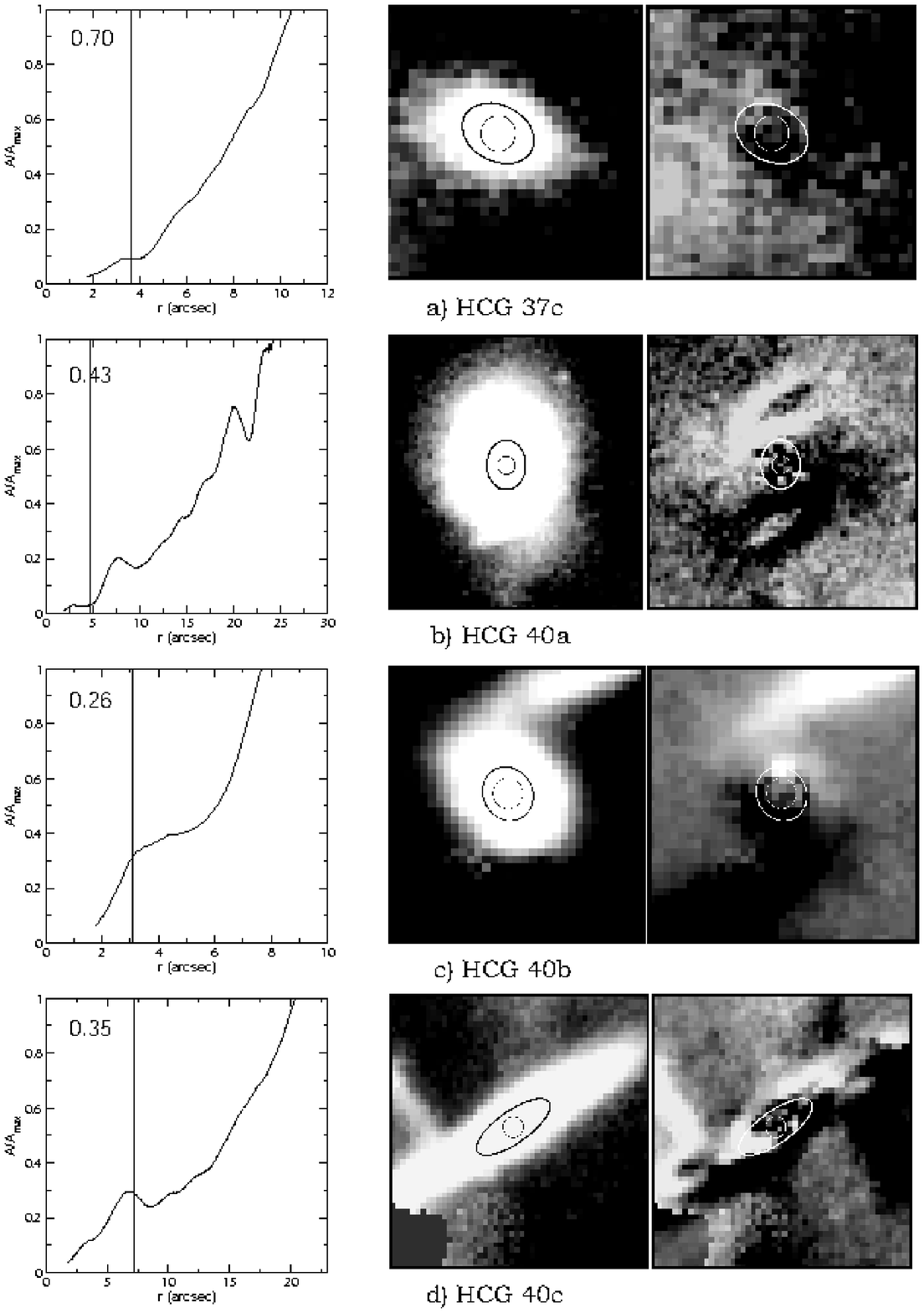} \caption{Same as
Figure~30.\label{fig31}}
\end{figure}
\clearpage

\begin{figure}
\epsscale{0.8}
\plotone{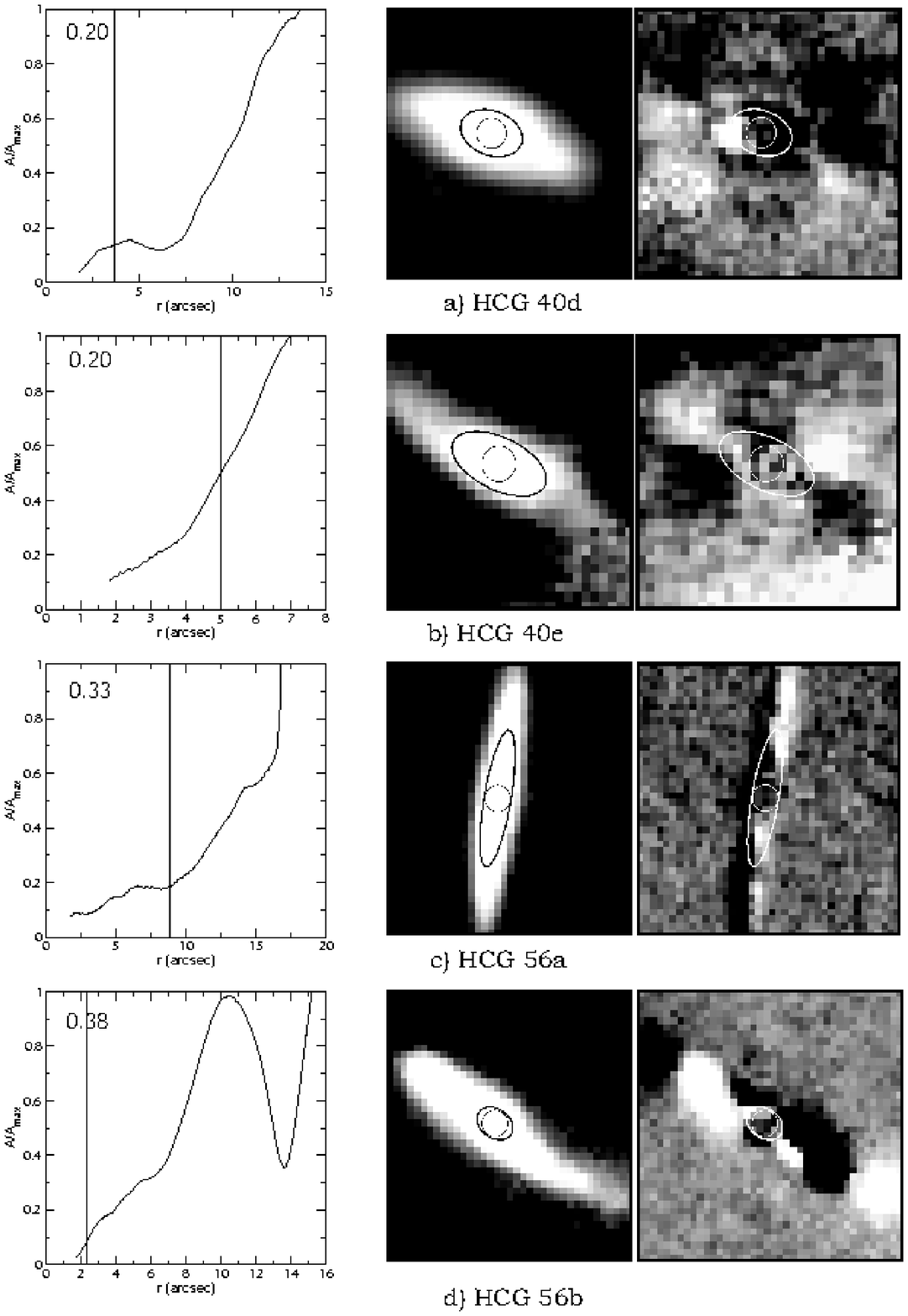} \caption{Same as
Figure~30.\label{fig32}}
\end{figure}
\clearpage

\begin{figure}
\epsscale{0.8}
\plotone{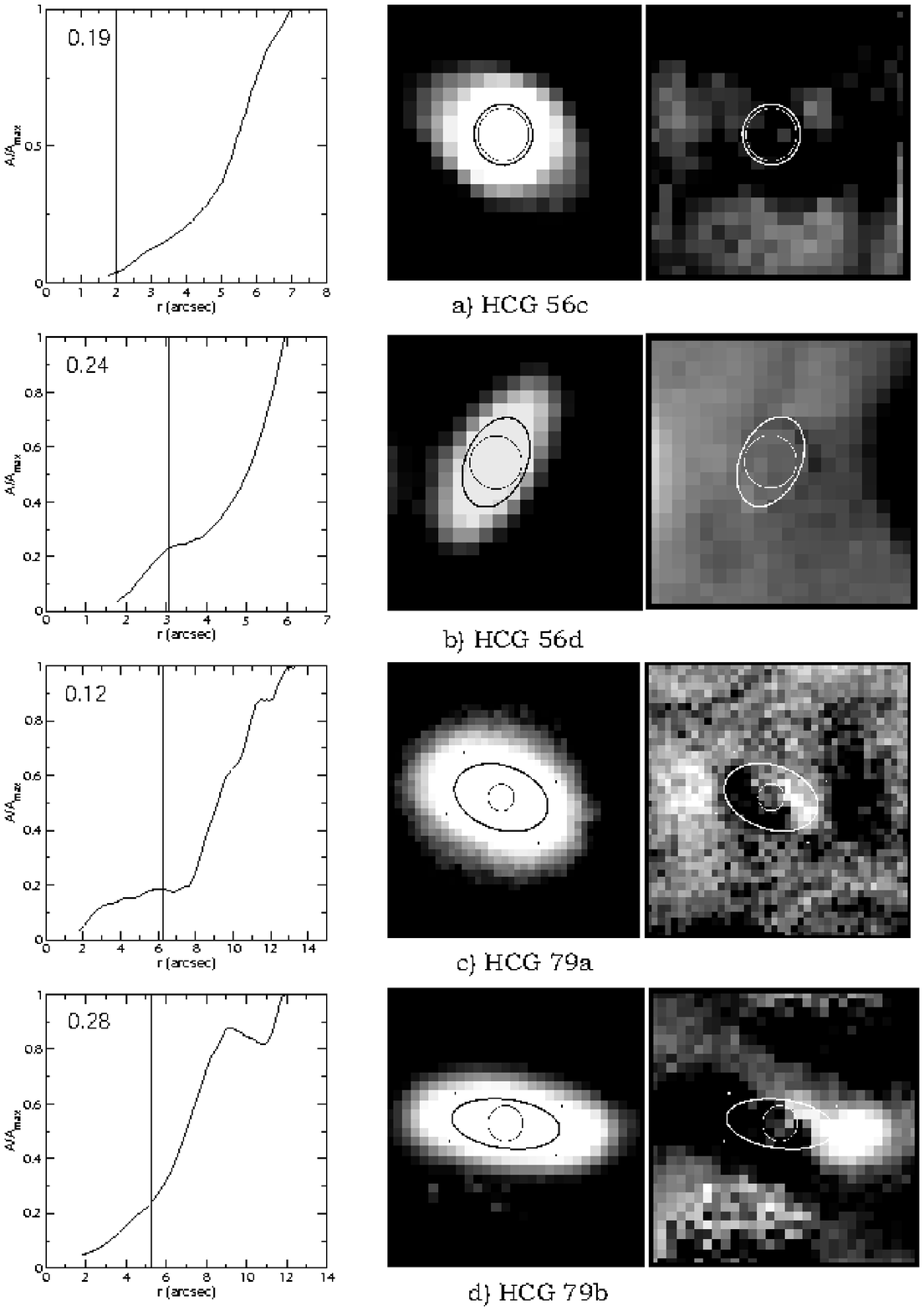} \caption{Same as
Figure~30.\label{fig33}}
\end{figure}
\clearpage

\begin{figure}
\epsscale{0.8}
\plotone{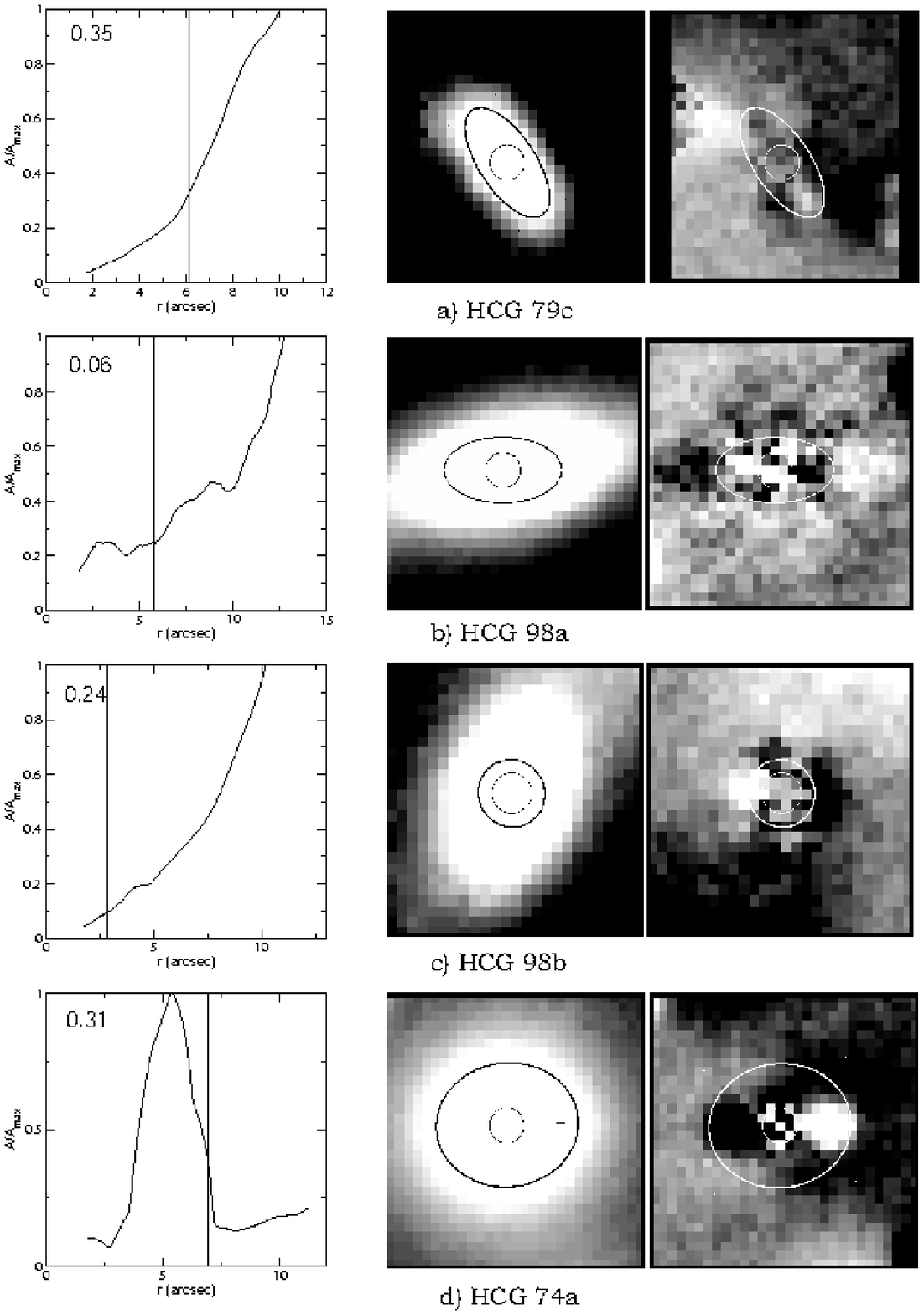} \caption{Same as
Figure~30.\label{fig34}}
\end{figure}
\clearpage

\begin{figure}
\epsscale{0.8} 
\plotone{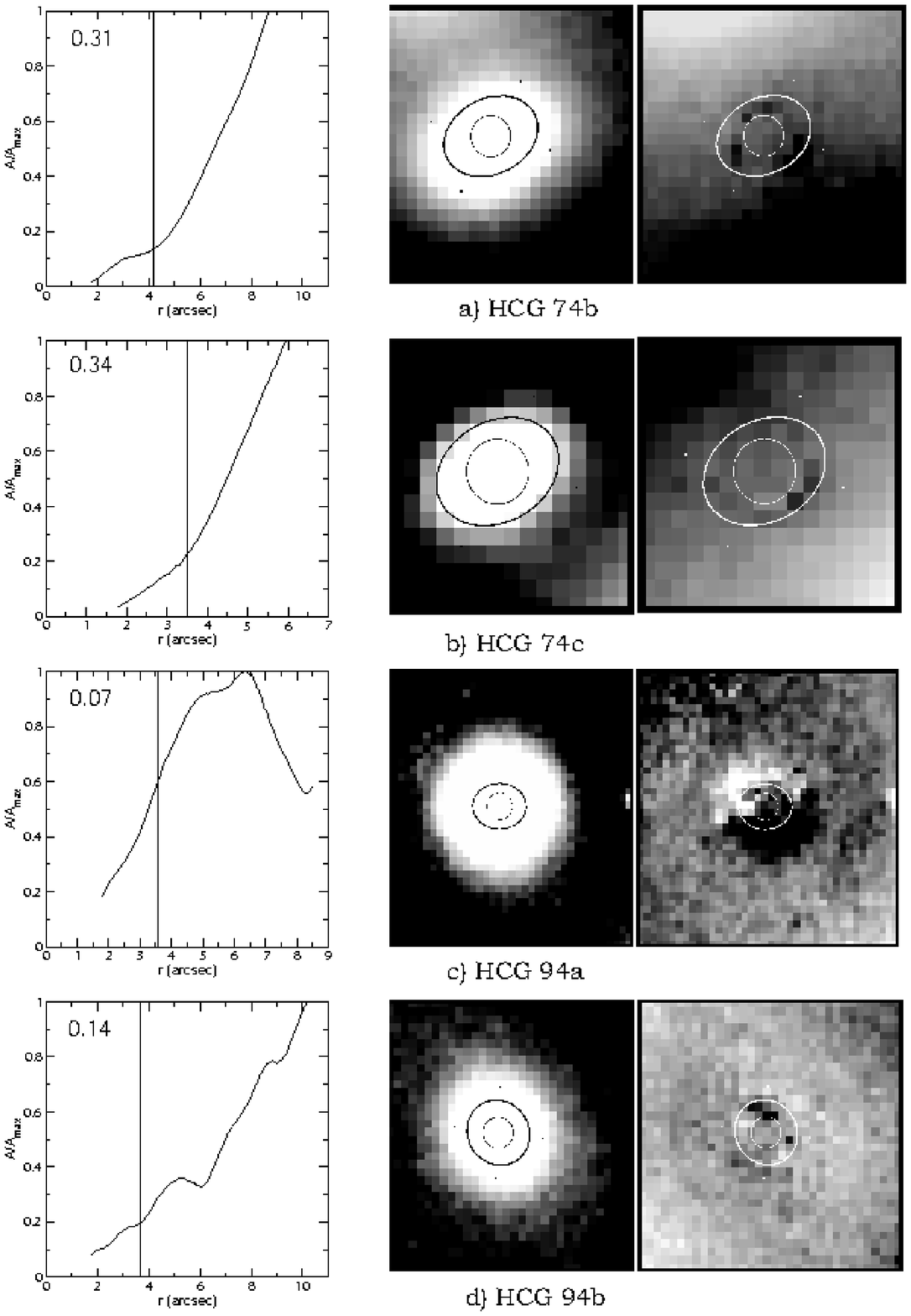} \caption{Same as
Figure~30.\label{fig35}}
\end{figure}
\clearpage

\begin{deluxetable}{lccccccccccc}
\tabletypesize{\scriptsize}
\tablecaption{Properties of the observed groups}
\tablewidth{0pt}
\tablehead{
\colhead{Group } & \colhead{ R.A.} & \colhead{ Dec.} &  \colhead{$\sigma $} & \colhead{$v_{o}$} &
\colhead{Angular size} & \colhead{Evolutionary} \\
\colhead{} &  \colhead{$(J2000)$} & \colhead{$(J2000)$} & \colhead{$(km\ s^{-1})$} & \colhead{$(km\ s^{-1})$} &
\colhead{$(arcmin)$} & \colhead{type} \\
\colhead{(1)} & \colhead{(2)} & \colhead{(3)} & \colhead{(4)} & \colhead{(5)} & \colhead{(6)} & \colhead{(7)}
}
\startdata
HCG~37 &$ 09\ 13\ 35.6$  & $+30\ 00\ 51$ & $692$  & $6685 $ & $3.2$ & B  \\
HCG~40 &$ 09\ 38\ 54.5$  & $-04\ 51\ 07$ & $251$  & $6685 $ & $1.7$ & B \\
HCG~56 &$ 11\ 32\ 39.6$  & $+52\ 56\ 25$ & $282$  & $8094 $ & $2.1$ & B \\
HCG~74 &$ 15\ 19\ 28.2$  & $+20\ 53\ 37$ & $537$  & $11962$ & $1.9$ & C \\
HCG~79 &$ 15\ 59\ 11.9$  & $+20\ 45\ 31$ & $229$  & $4347 $ & $1.9$ & B \\
HCG~88 &$ 20\ 52\ 22.8$  & $-05\ 45\ 28$ & $27 $  & $6026 $ & $5.2$ & A \\
HCG~94 &$ 23\ 17\ 16.5$  & $+18\ 43\ 11$ & $832$  & $12501$ & $2.8$ & C \\
HCG~98 &$ 23\ 54\ 12.7$  & $+00\ 22\ 24$ & $204$  & $7974 $ & $2.4$ & B \\
\enddata
\tablecomments{Columns are: (1) Hickson's group identifications
(Hickson 1982); (2) Right ascension (NASA/IPAC Extragalactic
Database (NED)); (3) Declination (NED); (4) Velocity dispersion of
the group (Hickson 1992); (5) Heliocentric group velocity (NED) ;
(6) Angular size subtended in the sky (NED); (7) Evolutionary type
(Coziol et al. 2004).}
\end{deluxetable}

\begin{deluxetable}{ccccccc}
\tabletypesize{\scriptsize}
\tablecaption{Properties of the observed galaxies}
\tablewidth{0pt} \tablehead{
\colhead{HCG} &  \colhead{Radial vel.} & \colhead{Morphology} & \colhead{$EW_{act}$} &
\colhead{Nuclear activity} &\colhead{$t_{J}$} & \colhead{$t_{K'}$}  \\
\colhead{} &  \colhead{($km\ s^{-1})$}  & \colhead{} & \colhead{(\AA)} & \colhead{} & \colhead{$(s)$} & \colhead{$(s)$} \\
\colhead{(1)} & \colhead{(2)} & \colhead{(3)} & \colhead{(4)} &
\colhead{(5)} & \colhead{(6)} & \colhead{(7)}}
\startdata
88b&$6010  $&  SBb  & $-6 $     & dSy2                      &$ 3600$  & $450      $ \\
88c&$6083  $&  Sc   & $-30$     & SFG                       &$ 3600$  & $200      $ \\
37a&$6745  $&  E7   & $-3 $     & dLINER                    &$ 2250$  & $860      $ \\
37b&$6741  $&  Sbc  & $-1 $     & LINER                     &$ 2250$  & $860      $ \\
37c&$7357  $&  S0a  & $\phn 0 $ & LLAGN?                    &$ 2250$  & $860      $ \\
40a&$6628  $&  E3   & $\phn 1 $ & no emission               &$ 1600$  & $900      $ \\
40b&$6842  $&  S0   & $\phn 3 $ & no emission               &$ 1600$  & $900      $ \\
40c&$6890  $&  Sbc  & $-12$     & SFG                       &$ 1600$  & $900      $ \\
40d&$6492  $&  SBa  & $-40$     & SFG                       &$ 1600$  & $900      $ \\
40e&$6625  $&  Sc   & $-3 $     & SFG?                      &$ 1600$  & $900      $ \\
56a&$8245  $&  Sc   & $-17$     & SFG                       &$ 3600$  & $900      $ \\
56b&$7919  $&  SB0  & $-52$     & Sy2                       &$ 3600$  & $900      $ \\
56c&$8110  $&  S0   & $\phn 1 $ & no emission               &$ 3600$  & $900      $ \\
56d&$8346  $&  S0   & $-24$     & SFG                       &$ 3600$  & $900      $ \\
56e&$7924  $&  S0   & $-14$     & SFG?                      &$ 3600$  & $900      $ \\
79a&$4292  $&  Sa   & $-5 $     & LINER                     &$ 3600$  & \nodata     \\
79b&$4446  $&  S0   & $-6 $     & SFG?                      &$ 3600$  & \nodata     \\
79c&$4146  $&  S0   & $\phn 5 $ & no emission               &$ 3600$  & \nodata     \\
98a&$7855  $&  SB0  & $\phn 9 $ & no emission               &$ 3600$  & $1440     $ \\
98b&$7959  $&  S0   & $\phn 3 $ & no emission               &$ 3600$  & $1440     $ \\
74a&$12255 $&  E1   & $\phn 1 $ & no emission               &$ 4400$  & $1160     $ \\
74b&$12110 $&  E3   & $\phn 2 $ & no emission               &$ 4400$  & $1160     $ \\
74c&$12266 $&  S0   & $\phn 3 $ & no emission               &$ 4400$  & $1160     $ \\
94a&$12040 $&  E1   & $\phn 5 $ & no emission               &$ 1300$  & $650      $ \\
94b&$11974 $&  E3   & $\phn 4 $ & no emission               &$ 1300$  & $650      $ \\
\enddata
\tablecomments{Columns are: (1) Galaxy identification (Hickson
1982); (2) Radial velocity (NED) (3) Morphological type (Hickson
1989), except for HCG~79a, which was taken from NED; (4, 5) Spectral
index and activity type (Coziol et al. 2004); (6) Total exposure
time in J; (7)  Total exposure time in K';}
\end{deluxetable}

 \begin{deluxetable}{lccccccccccc}
 \tabletypesize{\scriptsize}
 \tablecaption{Near Infrared magnitudes from \it 2MASS}
 \tablewidth{0pt}
 \tablehead{ \colhead{HCG} &  \colhead{$m_{J}\pm \Delta m_{J}$} &
 \colhead{$m_{H} \pm \Delta m_{H}$} &  \colhead{$m_{K}\pm \Delta m_{K}$}
& \colhead{$B-R$} \\
\colhead{} \\
 \colhead{(1)} &
 \colhead{(2)} &
 \colhead{(3)} &
 \colhead{(4)} &
 \colhead{(5)} }
 \startdata
  88b &  $11.04  \pm 0.04$    &  $10.38  \pm 0.04$     & $10.02  \pm  0.05$  & $1.51$  \\
  88c &  $12.44  \pm 0.04$    &  $11.77  \pm 0.05$     & $11.89  \pm  0.09$  & $0.99$  \\
  37a &  $10.25  \pm 0.02$    &  $9.60   \pm 0.02$     & $9.32   \pm  0.03$  & $1.79$  \\
  37b &  $11.60  \pm 0.03$    &  $10.88  \pm 0.03$     & $10.56  \pm  0.04$  & $1.99$  \\
  37c &  \nodata              &  \nodata               & \nodata             & $1.76$  \\
  40a &  $10.40  \pm 0.02$    &  $9.71   \pm 0.02$     & $9.46   \pm  0.03$  & $1.75$  \\
  40b &  $11.47  \pm 0.03$    &  $10.76  \pm 0.03$     & $10.52  \pm  0.04$  & $1.84$  \\
  40c &  $11.12  \pm 0.03$    &  $10.38  \pm 0.03$     & $10.12  \pm  0.05$  & $2.00$  \\
  40d &  $11.78  \pm 0.03$    &  $11.02  \pm 0.03$     & $10.67  \pm  0.04$  & $1.56$  \\
  40e &  $13.23  \pm 0.08$    &  $12.50  \pm 0.08$     & $12.18  \pm  0.08$  & $1.84$  \\
  56a &  $12.48  \pm 0.07$    &  $11.37  \pm 0.08$     & $10.48  \pm  0.05$  & $1.51$  \\
  56b &  $11.65  \pm 0.07$    &  $10.82  \pm 0.12$     & $10.01  \pm  0.06$  & $1.43$  \\
  56c &  \nodata              &  \nodata               & \nodata             & $1.52$  \\
  56d &  \nodata              &  \nodata               & \nodata             & $1.62$  \\
  56e &  $14.01  \pm 0.05$    &  $13.35  \pm 0.09$     & $12.97  \pm  0.08$  & $1.20$  \\
  79a &  $11.50  \pm 0.02$    &  $10.80  \pm 0.03$     & $10.53  \pm  0.04$  & $1.60$  \\
  79b &  $11.24  \pm 0.03$    &  $10.47  \pm 0.03$     & $10.59  \pm  0.06$  & $1.44$  \\
  79c &  \nodata              &  \nodata               & \nodata             & $1.27$  \\
  98a &  $10.40  \pm 0.07$    &  $9.67   \pm 0.09$     & $9.73   \pm  0.03$  & $1.60$  \\
  98b &  \nodata              &  \nodata               & \nodata             & $1.58$  \\
  74a &  $10.91  \pm 0.03$    &  $10.06  \pm 0.02$     & $9.85   \pm  0.04$  & $1.90$  \\
  74b &  \nodata              &  \nodata               & \nodata             & $1.88$  \\
  74c &  \nodata              &  \nodata               & \nodata             & $1.83$  \\
  94a &  $10.53  \pm 0.02$    &  $9.65   \pm 0.03$     & $9.60   \pm  0.04$  & $1.67$  \\
  94b &  \nodata              &  \nodata               & \nodata             & $1.63$  \\
 \enddata
 \tablecomments{Columns are: (1) HCG identification; (2) J magnitude;
 (3) H magnitude; (4) K magnitude; (5) B-R color (Hickson et al. 1989}
 \end{deluxetable}

\begin{deluxetable}{lccl}
\tabletypesize{\scriptsize} \tablecaption{Asymmetry analysis
summary} \tablewidth{0pt}
\tablehead{
\colhead{Evolutionary} & \colhead{HCG} &  \colhead{$r_{50}$}  & \colhead{asymmetric properties}  \\
\colhead{type}         &  \colhead{}   & \colhead{$(arcsec)$} & \colhead{} \\
\colhead{(1)} & \colhead{(2)} & \colhead{(3)} & \colhead{(4)}
}
\startdata
A      & 88b  & $9.3$ &  Tidal pair with merger -- blue nucleus\\
\phn   & +88c & $7.5$ &  Spiral arms -- blue nucleus  \\
B      & 37a  & $7.7$ &  Light excess toward b   \\
\phn   & 37b  & $10.$6&  Light excess toward a   \\
\phn   & 37c  & $3.7$ &  Symmetric -- blue nucleus\\
\phn   & 40a  & $4.7$ &  Tidal pair with bridge toward c -- blue nucleus? \\
\phn   & 40b  & $3.1$ &  Light excess toward c -- merger? -- blue nucleus?\\
\phn   & 40c  & $7.1$ &  Tidal pair -- bridges with e, b and a\\
\phn   & 40d  & $3.7$ &  Tidal pair with merger? -- light excesses toward a?\\
\phn   & 40e  & $5.0$ &  Tidal pair with bridge toward c -- blue nucleus \\
\phn   & -56a & $8.8$ &  Light excess toward other members\\
\phn   & -56b & $2.4$ &  Tidal pair with bridge toward c  \\
\phn   & 56c  & $2.0$ &  Symmetric.  \\
\phn   & 56d  & $3.1$ &  Symmetric.  \\
\phn   & 56e  & $2.6$ &  \nodata \\
\phn   & 79a  & $6.3$ &  Tidal pair with nucleus\\
\phn   & +79b & $5.3$ &  Bridge toward c -- tidal pair with tail  \\
\phn   & 79c  & $6.1$ &  Symmetric? -- bridge toward b \\
\phn   & +98a & $5.8$ &  Tidal pair with nucleus -- merger? -- blue nucleus? \\
\phn   & 98b  & $2.9$ &  Light excess toward 98a -- merger? \\
C      & 74a  & $7.0$ &  Tidal pair with merger\\
\phn   & 74b  & $4.1$ &  Symmetric \\
\phn   & 74c  & $3.5$ &  Symmetric \\
\phn   & +94a & $3.6$ &  Light excess -- merger? -- blue nucleus\\
\phn   & 94b  & $3.7$ &  Light excess -- blue nucleus\\
\enddata
\tablecomments{Columns are: (1) Global activity type; (2) HCG
identification (a + sign identifies galaxies with blue 2MASS colors and a - sign identifies
galaxies affected by AGN or star formation with dust extinction); (3) Radius containing 50\% of total light; (4) Observed asymmetric properties.}
\end{deluxetable}

\begin{deluxetable}{lccc}
\tabletypesize{\scriptsize}
\tablecaption{Asymmetry vs. activity analysis}
\tablewidth{0pt}
\tablehead{
\colhead{} & \multicolumn{3}{c}{Evolutionary type}\\
\colhead{} & \colhead{A} & \colhead{B}& \colhead{C}}
\startdata
Nb. of gal.  &2 & 18 & 5 \\
Nb. active       &2 & 12 & 0 \\
Nb. late-type    &2 & 5  & 0 \\
Nb. asymmetric    &1 & 13 & 3 \\
\enddata
\end{deluxetable}


\clearpage

\begin{thebibliography}{}
\bibitem[]{1032} Abraham, R. G., Valdes, F., Yee, H. K. C. \& van den Bergh, S. 1994, \apj, 432, 75
\bibitem[]{1033} Abraham, R. G., Tanvir, N. R., Santiago, B. X., Ellis, R. S., Glazebrook, K. \& van den Bergh, S. 1996, \mnras, 279, L47
\bibitem[]{1034} Allam, S. S., Tucker, D. L., Lin, H. \& Hashimoto, Y. 1999, \apj, 522, L89
\bibitem[]{1035} Andernach, H. \& Coziol, R. 2006, Proc. ESO Workshop: Groups of galaxies
in the nearby Universe, ESO Astrophysics Symposia, I. Saviane, V.
Ivanov, J. Borissova, eds., Springer-Verlag, 2006, in press (astro-ph/0603295)
\bibitem[]{1038b} Barth, C. S., Coziol, R. \& Demers, S. 1995, \mnras, 276, 1224
\bibitem[]{1039} Belokurov, V., Zucker, D. B., Evans, N. W., Wilkinson, M. I. and 29 coauthors 2006, \apj, 647, 111
\bibitem[]{1040} Bender, R., \& M\"ollenhoff, C. 1987, \aap, 177, 71
\bibitem[]{1041} Bender, R., Burstein, D. \& Faber, S. M. 1993, \apj, 411, 153
\bibitem[]{1042} Byrd, G. \& Valtonen, M.  1990, \apj, 350, 89
\bibitem[]{1043} Caon, N., Capaccioli, M., D'Onofrio, M., \& Longo, G. 1994, \aap, 286, 39
\bibitem[]{1044} Conselice, C. J. 1997, \pasp, 109, 1251
\bibitem[]{1045} Conselice, C. J. \& Gallagher, J. S. 1999, \aj, 117, 75
\bibitem[]{1046} Conselice, C. J. \& Bershady, M. A. 2000, \apj, 529, 886
\bibitem[]{1047} Coziol, R., Ribeiro, A. L. B., Capelato, H. V. \& de~Carvalho, R. R. 1998a, \apj, 493, 563
\bibitem[]{1048} Coziol, R., de~Carvalho, R. R., Capelato, H. V. \& Ribeiro, A. L. B. 1998b, \apj, 506, 545
\bibitem[]{1049} Coziol, R., Iovino, A. \& de~Carvalho, R. R. 2000, \aj, 120, 47
\bibitem[]{1050} Coziol, R., Brinks, E., \& Bravo-Alfaro, H. 2004, \aj, 128, 68
\bibitem[]{1051} Cruz-Gonz\'alez, I., Carrasco, L., Ruiz, E., Salas, L., Skrutskie, M., Meyer, M., Sotelo, P.,
Barbosa, F., Guti\'erez, L., Iriarte, A., Cobos, F., Bernal, A., S\'anchez, B., Vald\'ez, J.,
Arg\"uelles, S., Conconi, P. 1994, in Instrumentation in Astronomy
VIII, D. L. Crawford \& E. R. Craine, eds. Proc. SPIE 2198, 774
\bibitem[]{1055} Da Rocha, C. \& Mendes de Oliveira, C. 2005, \mnras, 364, 1069
\bibitem[]{1056} de la Rosa, I. G., de Carvalho, R. R., Vazdekis, A. \& Barbuy, B 2006, RevMexAA (Serie de Conferencias), 26, 113
\bibitem[]{1057} Di Tullio, G. A. 1979, \aaps, 37, 591
\bibitem[]{1058} Dressler, A. 1980 \apj, 236, 315
\bibitem[]{1059} Ellingson, E. 2003, in Galaxy Evolution in Groups and Clusters, C. L. Lobo, M. S. Roos, A. Biviano, eds. \apss, 285, 9.
\bibitem[]{1060} Frogel, J. A., Persson, S. E., Aaronson, M. \& Matthews, K. 1978, \apj, 220, 75
\bibitem[]{1061} Frogel, J. A. \& Elias, J. H. 1987, \apj, 313, L53
\bibitem[]{1062} Fujita, Y. 1998, \apj, 509, 587
\bibitem[]{1063} G\'omez-Flechoso \& M. A., Dom\'{\i}nguez-Tenreiro, R. 2001, \apj, 549, L187
\bibitem[]{1064} Gunn, J. E. \& Gott, J. R. 1972, \apj, 176,1
\bibitem[]{1065} Henriksen, M. \& Byrd, G. 1996, \apj, 459, 82
\bibitem[]{1066} Hibbard, J. E. \& Mihos, J. C. 1995 AJ, 110, 140
\bibitem[]{1067} Hickson, P. 1982, \apj, 255, 382
\bibitem[]{1068} Hickson, P., Kindl, E. \& Huchra J. P. 1988, \apj, 331, 64
\bibitem[]{1069} Hickson, P., Kindl, E. \& Auman, J. R. 1989, \apjs, 70, 687
\bibitem[]{1070} Hickson, P., Mendes de Oliveira, C., Huchra, J.~P., \& Palumbo, G.~G. 1992, \apj, 399, 353
\bibitem[]{1071} Hutchmeier, W. K. 1997, \aap, 325, 473
\bibitem[]{1072} Jarret, T. H., Chester, T., Cutri, R., Schneider, S. E. \& Huchra, J. P. 2003, \aj, 125, 525
\bibitem[]{1073} King, I. R. 1977, in The Evolution of Galaxies and Stellar
Populations, B. Tinsley \& R. Larson eds., New Haven: Yale Univ. Obs., p. 418
\bibitem[]{1073b} Ko, J. \& Im, M. 2005, Journal of the Korean Astronomical Society, 39, 3
\bibitem[]{1075} Kotilainen, J. K. \& Ward, M. J. 1994, \mnras, 266, 953
\bibitem[]{1076} Kormendy, J. 1982, in Morphology and Dynamics of Galaxies, Martinet L., Mayor M., eds., Saas Fee, p. 113
\bibitem[]{1076b} La Barbera, F., Busarello, G., Massarotti, M., Merluzzi, P. \& Mercurio A.  2003, \aap, 409, 21
\bibitem[]{1077} Leon, S., Combes, F. \& Menon, T. K. 1998, \aap, 330, 37
\bibitem[]{1078} Martin, N. F., Ibata, R. A., Bellazzini, M., Irwin, M. J., Lewis, G. F. \& Dehnen, W. 2004, \mnras, 348, 12
\bibitem[]{1079} Martin, N. F., Ibata, R. A., Irwin, M. J., Chapman, S., Lewis, G. F., Ferguson, A. M. N.,
Tanvir, N. \& McConnachie, A. W. 2006, MNRAS, 947, in press (arXiv:astro-ph/0607472)
\bibitem[]{1081} McConnachie, A. W. \& Irwin, M. J. 2006, \mnras, 365, 1263
\bibitem[]{1082} Mendes de Oliveira, C. \& Hickson, P. 1994, \apj, 427, 684
\bibitem[]{1083} Merritt, D. 1984, \apj, 276, 26
\bibitem[]{1084} Michard, R. 1999, \aaps, 137, 245
\bibitem[]{1085} Michard, R. 2000, \aap, 360, 85
\bibitem[]{1086} Nishiura, S., Murayama, T., Shimada, M., Sato, Y., Nagao, T., Molikawa, K., Taniguchi, Y. \& Sanders, D. B.
2000, \aj, 120, 2355
\bibitem[]{1088} Mihos, J. C. \& Hernquist, L. 1996 \apj, 464, 641
\bibitem[]{1089} Mihos, J. C. 2004, in Clusters of Galaxies: Probes of Cosmological Structure and Galaxy Evolution,
Carnegie Observatories Astrophysics Series, J.S. Mulchaey, A.
Dressler, and A. Oemler, eds. Cambridge University Press, p. 277
\bibitem[]{1092} Moore, B., Katz, N., Lake, G., Dressler, A., Oemler, A., Jr.  1996, Nature, 379, 613
\bibitem[]{1093} Moore, B., Lake, G., \& Katz N. 1998, \apj, 495, 139
\bibitem[]{1093b} Naab, T., Khochfar, S. \& Burkert, A. 2006a,\apj, 636,L81
\bibitem[]{1093c} Naab, T., Jesseit, R. \& Burkert, A. 2006b, \mnras, 372, 839
\bibitem[]{1094} Oemler, A. 1974, \apj, 194, 1
\bibitem[]{1095} Pildis, R. A., Bregman, J. N. \& Schombert, J. M. 1995, \aj, 110, 1498
\bibitem[]{1096} Sadler, E. M., 1984, \aj, 89, 23
\bibitem[]{1096b} Schweizer, F. \& Seitzer, P. 1992, \aj, 104, 1039
\bibitem[]{1097} Tapia, M., Neri, L. \& Roth, M. 1986, RMxAA, 13, 115
\bibitem[]{1097b} Hern\'andez-Toledo, H. M., Avila-Reese, V., Salazar-Contreras, J. R. \& Conselice, C. J. 2006, \aj, 132, 71
\bibitem[]{1097c} Toomre, A. 1977 ARA\&A, 15, 437
\bibitem[]{1098} van Dokkum, P. G. 2005, \aj, 130, 2647
\bibitem[]{1099} Verdes-Montenegro, L., Yun, M. S., Williams, B. A., Huchtmeier, W. K., Del Olmo, A. \& Perea, J. 2001, \aap, 377, 812
\bibitem[]{1100} Valageas, P., Schaeffer, R. \& Silk, J. 2003, \mnras, 344, 53
\bibitem[]{1101} Yoshioka, T., Furuzawa, A., Takahashi, S., Tawara, Y., Sato, S. Yamashita, K. \& Kumai, Y.
2004, Advances in Space Reasearch, 34, 2525
\bibitem[]{1103} York, D. G., Adelman, J., Anderson, J. E., Anderson, S. F., Annis, J.,
Bahcall, N. A., Bakken, J. A., Barkhouser, R., Bastian, S., Berman, E., \& 134 coautores 2000, \aj, 120, 1579
\bibitem[]{1105} Zaritsky, D. \& Lo, K. Y. 1986, \apj, 303, 66
\bibitem[]{1106} Zepf, S. E., Whitmore, B. C. \& Levison, H. F. 1991, \apj, 383, 524
\bibitem[]{1106b} Zepf, S. E. \& Whitmore, B. C. 1991, \apj, 383, 542
\bibitem[]{1107} Zucker, D. B., Belokurov, V., Evans, N. W., Wilkinson, M. I. and 29 coauthors 2006, \apj, 643, 103
\end{thebibliography}
\end{document}